# An algorithm to assign musical prime commas to every prime number and construct a universal and compact free Just Intonation musical notation

**Dr David Ryan, Edinburgh, UK**

**Draft 05, March 2017**

## 1) Abstract

Musical frequencies in Just Intonation are comprised of rational numbers. The structure of rational numbers is determined by prime factorisations. Just Intonation frequencies can be split into two components. The larger component uses only integer powers of the first two primes, 2 and 3. The smaller component decomposes into a series of microtonal adjustments, one for each prime number 5 and above present in the original frequency. The larger 3-limit component can be notated using scientific pitch notation modified to use Pythagorean tuning. The microtonal adjustments can be notated using rational commas which are built up from prime commas. This gives a notation system for the whole of free-JI, called Rational Comma Notation. RCN is compact since all microtonal adjustments can be represented by a single notational unit based on a rational number. RCN has different versions depending on the choice of algorithm to assign a prime comma to each prime number. Two existing algorithms SAG and KG2 are found in the literature. A novel algorithm DR is developed based on discussion of mathematical and musical criteria for algorithm design. Results for DR are presented for primes below 1400. Some observations are made about these results and their applications, including shorthand notation and pitch class lattices. Results for DR are compared with those for SAG and KG2. Translation is possible between any two free-JI notations and any two versions of RCN since they all represent the same underlying set of rational numbers.

## 2) Introduction to free-JI notation systems

Just Intonation (JI) is the study of musical harmony where the frequency ratios between musical notes take whole numbered values. Every note in a JI musical composition has a frequency which is a rational-numbered multiple of a suitable reference frequency. Relative frequencies (hereafter, just frequencies) in JI can thus be represented by rational numbers, such as 1/1, 2/1, 5/4, 4/7, 13/11, 80/81, 65537/2. Note that the normal form for rational numbers is to be 'reduced' which is to have no common factors on top and bottom, so 4/2 reduces to 2/1, 15/6 reduces to 5/2, etc.

In mathematics, the structure of rational numbers is well understood, and the fundamental theorem of arithmetic states that every whole number has a unique factorisation in terms of prime numbers raised to non-negative powers. Moreover, every rational number has a unique factorisation in terms of prime numbers raised to any integer powers. The sequence of prime numbers starts 2, 3, 5, 7, 11, 13, 17, 19, 23, 29… (OEIS 2016, sequence 'A000040').

JI can be restricted to only use frequencies with the smallest few primes as factors (up to 2, 3, 5, 7…); these systems are called '$p$-limit JI' and use only the '$p$-smooth' whole numbers:

- 2-limit JI contains only a set of octaves relative to the reference frequency (…1/4, 1/2, 1/1, 2/1, 4/1, 8/1…) which is of limited musical use due to the large octave-sized gaps between successive notes.
- 3-limit JI contains the 'Pythagorean' frequencies such as 1/1, 2/1, 3/2, 4/3, 9/8. There is a 3-limit number arbitrarily close to any rational number, so it is the smallest $p$-limit which is dense on the number line. This will make it useful for constructing commas for each prime number.
- 5-limit JI contains much of classical harmony, such as major/minor thirds (5/4, 6/5), major/minor sixths (5/3, 8/5) as well as useful semitones (16/15, 25/24) and commas (80/81).
- 7-limit JI contains much of extended classical and jazz harmony such as seventh chords: the dominant seventh interval 7/4 is very useful, as well as subminor third 7/6, also 8/7, 21/16, comma 63/64.
- Higher $p$-limits for $p$ = 11, 13… have not been used extensively (to date) in musical composition. Note – some authors prefer 'Rational Intonation' instead of 'Just Intonation' when the $p$-limit is high, say above 20 to 50 depending on the author, for example '2477-limit Rational Intonation' includes whole numbered frequencies up to 2500. However, the term 'Just Intonation' will be retained here for higher *p*-limits, e.g. '2477-limit JI'.

- Free-JI is just the same as the whole of JI, but emphasising the lack of a prime limit.

One way of accessing free-JI is to compose directly in whole-numbered Hertz; some higher primes may appear as factors of the whole numbered frequencies, e.g. a short melody 200Hz, 250Hz, 257Hz contains a high prime factor 257. Another way is to use a suitably powerful notation comprising of traditional note names with suitable modifiers. This paper presents a notation (RCN) of that form, and also reviews how well several other notation systems in the literature cope with free-JI.

One obstacle a good free-JI notation must overcome is that there are an infinite number of primes, each of which can occur in any positive or negative power in any given frequency. Providing for all these infinite options makes it difficult to construct a JI notation system which is simultaneously precise, concise, and readable.

Fortunately, this construction is aided by JI frequencies being decomposed into two components:

- A Pythagorean component in 3-limit JI which is usually within a semitone of the original note
- A series of microtonal adjustments from this Pythagorean approximation to the desired rational frequency

Subsequently, a JI notation can be based on combining a suitable 3-limit notation with a suitable notation for the microtonal adjustments. This is the basis for Rational Comma Notation (RCN) which is defined below.

## 3) Overview of modified SPN as a 3-limit notation

The 3-limit notation requires two components to specify an exact frequency. Although these could be the numbers $(a, b)$ in $2^a3^b$, a more usable pair of components are $(\mathrm{L}, z)$ or $\mathrm{L}_z$ where $z$ is an octave number and L is a pitch class label. This style of notation is used for Scientific Pitch Notation (SPN) which assigns notations to 12-EDO notes. Examples of SPN include: $C_4$ for middle C, $C_5$ for the next higher C, $A_4$ for concert A, $E_7$ is a high E, $F\#_3$ is the F# below middle C, $Bb_{-53}$ is an astrophysically low note sung by a black hole. Note that the musical sharp and flat symbols will be typed as '#', 'b' respectively, and that the flat symbol 'b' is distinct from the note label 'B'.

It is possible to reassign SPN pitch class labels to use Pythagorean (3-limit) pitch classes. Examples of assigning SPN notations to Pythagorean notes include: assign 1/1 to $C_4$, this implies 2/1 is $C_5$, assign 3/2 to $G_4$. assign 27/16 to $A_4$, assign 81/64 to $E_4$, this implies 81/8 is $E_7$, assign 16/9 to $Bb_4$, assign 2187/2048 to $C\#_4$, etc. In this modified SPN, sharps and flats no longer cancel out and no enharmonic note labels exist – for example, C# will be a different pitch class to Db, and F### is not the same pitch class as G#.

Note that the assignment of 1/1 to $C_4$ is arbitrary, although it is the convention followed throughout this paper. When working in a key signature other than C it is common to assign 1/1 to the note label of that key signature, e.g. when working in D then 1/1 could be assigned to $D_4$ and then $C_4$ would become 8/9. Because of this, some authors speak instead in terms of 'Pythagorean offset' which is the 3-exponent of the 3-limit number. For example, 1/1, 3/1, 81/64, 16/9 have Pythagorean offsets 0, 1, 4, -2 respectively.

In this paper the convention will be to speak of 1/1 as a 'C' note due to the simplicity and intuitiveness of this way of speaking. This means that …F, C, G, D… are being used here as a code for Pythagorean offsets …-1, 0, 1, 2…. So the ideas in this paper could be restated in terms of Pythagorean offsets, and thus apply to every key signature, not just C. Where concepts are already stated in terms of 3-exponent, that number is not dependent on C being 1/1 and so the concept applies in every key.

**Table 1: Pythagorean notes in octave 4 with frequencies in fractional, decimal and cents forms**

| Note | Fract. | Dec. | Cents | Note | Fract. | Dec. | Cents | Note | Fract. | Dec. | Cents |
|---|---|---|---|---|---|---|---|---|---|---|---|
| $Fb_4$ | 8192 /6561 | 1.2486 | 384.36 | $F_4$ | 4/3 | 1.3333 | 498.04 | $F\#_4$ | 729/512 | 1.4238 | 611.73 |
| $Cb_4$ | 2048 /2187 | 0.9364 | -113.69 | $C_4$ | 1/1 | 1.0000 | 0.00 | $C\#_4$ | 2187 /2048 | 1.0679 | 113.69 |
| $Gb_4$ | 1024 /729 | 1.4047 | 588.27 | $G_4$ | 3/2 | 1.5000 | 701.96 | $G\#_4$ | 6561 /4096 | 1.6018 | 815.64 |
| $Db_4$ | 256 /243 | 1.0535 | 90.22 | $D_4$ | 9/8 | 1.1250 | 203.91 | $D\#_4$ | 19683 /16384 | 1.2014 | 317.60 |
| $Ab_4$ | 128 /81 | 1.5802 | 792.18 | $A_4$ | 27/16 | 1.6875 | 905.87 | $A\#_4$ | 59049 /32768 | 1.8020 | 1019.55 |
| $Eb_4$ | 32/27 | 1.1852 | 294.13 | $E_4$ | 81/64 | 1.2656 | 407.82 | $E\#_4$ | 177147 /131072 | 1.3515 | 521.51 |
| $Bb_4$ | 16/9 | 1.7778 | 996.09 | $B_4$ | 243 /128 | 1.8984 | 1109.78 | $B\#_4$ | 531441 /262144 | 2.0273 | 1223.46 |

With 1/1 assigned to $C_4$, a total of 21 notes from the resulting 3-limit notation scheme are described above in Table 1. The seven traditional note labels 'F, C, G, D, A, E, B' are in columns which are grouped by shading, with the flatted, natural and sharped variants in rows. Double flats, double sharps (etc.) would extend this table to the left and right. Note that $Cb_4$ and $C_4$ must be 113 cents apart, the same for $B_4$ and $B\#_4$, so these stray slightly outside the expected range for octave 4 which is 0 to 1200 cents; this little-known and paradoxical feature was also part of the original SPN scheme for 12-EDO.

**Table 2: Pythagorean notes between 0 and 1200 cents with at most one sharp or flat**

| **Note** | $C_4$ | $B\#_3$ | $Db_4$ | $C\#_4$ | $D_4$ | $Eb_4$ | $D\#_4$ | $Fb_4$ | $E_4$ | $F_4$ | $E\#_4$ |
|---|---|---|---|---|---|---|---|---|---|---|---|
| **Fraction** | 1/1 | 531441 /524288 | 256 /243 | 2187 /2048 | 9/8 | 32/27 | 19683 /16384 | 8192 /6561 | 81/64 | 4/3 | 177147 /131072 |
| **Cents** | 0.0 | 23.5 | 90.2 | 113.7 | 203.9 | 294.1 | 317.6 | 384.4 | 407.8 | 498.0 | 521.5 |
| **Note** | $Gb_4$ | $F\#_4$ | $G_4$ | $Ab_4$ | $G\#_4$ | $A_4$ | $Bb_4$ | $A\#_4$ | $Cb_5$ | $B_4$ | $C_5$ |
| **Fraction** | 1024 /729 | 729 /512 | 3/2 | 128/81 | 6561 /4096 | 27/16 | 16/9 | 59049 /32768 | 4096 /2187 | 243 /128 | 2/1 |
| **Cents** | 588.3 | 611.7 | 702.0 | 792.2 | 815.6 | 905.9 | 996.1 | 1019.6 | 1086.3 | 1109.8 | 1200.0 |

The same 21 pitch classes have been illustrated in Table 2 but shifted into the range 0 to 1200 cents. All notes are in octave 4, except B#$_3$, Cb$_5$ and C$_5$. The notes have been reordered into ascending order of frequency. One observation is that with 21 pitch classes the pitch class labels become jumbled – for example, an ascending sequence of Eb, D#, Fb, E occurs. This jumbling, or naming paradox, happens whenever there are more than 12 pitch classes in a Pythagorean scale, since after sorting by ascending frequency there will be sequences of note labels such as (Db, C#) where the note labels (ignoring sharps and flats) are in descending order.

This 3-limit notation scheme suffers from a problem that the sequences B, B#, B##, B###... and C, Cb, Cbb, Cbbb… have octave numbers which travel outside of the octave. One possible remedy is to factor out Pythagorean commas. A Pythagorean comma is the note B#$_3$ = 531441/524288 (see Table 2) which is a small 'comma' interval of 23.46 cents. An inverse Pythagorean comma is Dbb$_4$ = 524288/531441, -23.46 cents. Just as sharps '#' and flats 'b' are shorthand for 2187/2048 and 2084/2187 respectively, Pythagorean commas B#$_3$ and inverse Pythagorean commas Dbb$_4$ can be given shorthands, say p = B#$_3$ and d = Dbb$_4$, leading to alternative names for some Pythagorean intervals:

**Table 3: Alternative Pythagorean note names, using p and d for forward and inverse Pythagorean commas**

| **Note name** | C$_4$ | B#$_3$ | Db$_4$ | C#$_4$ | D$_4$ | Eb$_4$ | D#$_4$ | Fb$_4$ | E$_4$ | F$_4$ | E#$_4$ |
|---|---|---|---|---|---|---|---|---|---|---|---|
| **Alt. name** | B#d$_3$ | Cp$_4$ | C#d$_4$ | Dbp$_4$ | | D#d$_4$ | Ebp$_4$ | Ed$_4$ | Fbp$_4$ | E#d$_4$ | Fp$_4$ |
| **Cents** | 0.0 | 23.5 | 90.2 | 113.7 | 203.9 | 294.1 | 317.6 | 384.4 | 407.8 | 498.0 | 521.5 |
| **Note name** | Gb$_4$ | F#$_4$ | G$_4$ | Ab$_4$ | G#$_4$ | A$_4$ | Bb$_4$ | A#$_4$ | Cb$_5$ | B$_4$ | C$_5$ |
| **Alt. name** | F#d$_4$ | Gbp$_4$ | | G#d$_4$ | Abp$_4$ | | A#d$_4$ | Bbp$_4$ | Bd$_4$ | Cbp$_5$ | |
| **Cents** | 588.3 | 611.7 | 702.0 | 792.2 | 815.6 | 905.9 | 996.1 | 1019.6 | 1086.3 | 1109.8 | 1200.0 |

In Table 3 above the shorthands 'p' and 'd' have been used to rewrite the notes in Table 2 in terms of only twelve consecutive note labels (in this case Db, Ab, Eb, Bb, F, C, G, D, A, E, B, F# in black text) with labels outside this range in grey text. Using 'p' and 'd', many of these notes form pairs separated by a Pythagorean comma. The notations in each pair transform into each other by adding 'p' or 'd' to the notation. Also the octave number problem is partially solved, since B#$_3$ is now Cp$_4$, and Cb$_5$ is now Bd$_4$, so both of these have a note name in octave 4. There will still be some notations which 'travel outside' of the octave such as Cd$_4$ and Bpppp$_4$, however these are outside by much smaller distances than Cb$_4$ and B#$_4$ respectively from Table 2.

Using 'p' and 'd' does not shorten the more remote Pythagorean notations very much; for example, 12 copies of '#' have the same pitch class as 7 copies of 'p'; A### is equivalent to Cpp, G#### is equivalent to Cppp, A####### is equivalent to Fppppp, Dbbbbb is equivalent to Addd. This is because '#' and 'p' give a 3-exponent of 7 and 12 respectively. In order to make the remote Pythagorean notations significantly shorter, a shorthand for a higher power of 3 could be used. Good commas to approximate might be $3^{53}2^{-84}$ at +3.615 cents, or $3^{665}2^{-1054}$ at +0.07558 cents. These 3-exponents 53 and 665 are denominators of convergents for $\log_2 3$ expressed as a continued fraction (see OEIS 2016, 'sequence A005664') which have particularly large jumps to the next denominator in the series, meaning that 3 to the power of one of these denominators provides an excellent approximation to a whole number of octaves. Assigning extra symbols beyond ‘p’ and ‘d’ remain as further work.

**Table 4: Powers of 3 from 1 to $3^{24}$ and their note labels (some alternative names given in brackets)**

| **3-exponent** | 0 | 1 | 2 | 3 | 4 | 5 | 6 | 7 | 8 | 9 | 10 | 11 | 12 |
|---|---|---|---|---|---|---|---|---|---|---|---|---|---|
| **Integer** | 1 | 3 | 9 | 27 | 81 | 243 | 729 | 2,187 | 6,561 | 19,683 | 59,049 | 177,147 | 531,441 |
| **Note Label** | C | G | D | A | E | B | F# | C# | G# | D# | A# | E# | B# (Cp) |

| **3-exp.** | 13 | 14 | 15 | 16 | 17 | 18 | 19 |
|---|---|---|---|---|---|---|---|
| **Integer** | 1,594,323 | 4,782,969 | 14,348,907 | 43,046,721 | 129,140,163 | 387,420,489 | 1,162,261,467 |
| **Label** | F## | C## | G## | D## | A## (Bp) | E## | B## (C#p) |

| **3-exp.** | 20 | 21 | 22 | 23 | 24 |
|---|---|---|---|---|---|
| **Integer** | 3,486,784,401 | 10,460,353,203 | 31,381,059,609 | 94,143,178,827 | 282,429,536,481 |
| **Label** | F### (G#p) | C### | G### | D### | A### (Cpp) |

The integer powers of 3 can be denoted $3^k$ for $k \geq 0$. These numbers are part of both the overtone series and 3-limit JI. Each represents the first time a new Pythagorean note label (unaccompanied by any prime commas) is used in the integers. The integer powers are given in Table 4 above for $0 \leq k \leq 24$. It can be deduced that all Pythagorean integers below 1,594,323 require only a single sharp, and that A### = Cpp first appears at around 282 billion.

## 4) Overview of rational commas as a microtonal adjustment scheme

Now the microtonal adjustment scheme will be described. It needs a few simple properties to work well:

- For each prime number 5 and above, the prime $p$ has a uniquely chosen frequency adjustment (a ‘prime comma’) of the form $[p] = 2^a 3^b p$. (In what follows, let ‘higher prime’ be synonymous with ‘prime 5 and above’.)
    - Each prime comma $[p]$ is a rational number, a fraction, a small shift of frequency up or down
    - Each prime comma $[p]$ is microtonal, i.e. less than a semitone in size
    - Each prime comma $[p]$ pairs a prime number $p$ with a Pythagorean (rational, 3-limit) number $2^{-a}3^{-b}$ of approximately the same size
    - $2^{-a}3^{-b}$ may be an integer ($a$, $b \leq 0$), especially for large $p$, but for small $p$ it is usually a fraction
    - The 3-exponent of the Pythagorean ($-b$) is the opposite sign to that of the comma ($b$)
- A prime comma adjustment can be reversed by taking the reciprocal: $[1/p] = 1/[p]$
- Prime commas can be multiplied or divided to define rational commas $[pq]$ and $[p/q]$: $[p][q] = [pq]$ and $[p]/[q] = [p/q]$ which is the essence of the ‘compact’ property defined below
- Given any $x$, $y$ with only prime factors 5 and above, i.e. the 5-rough numbers, $[xy]$ and $[x/y]$ can thus be derived iteratively from the prime factors of $x$ and $y$

- The identity comma [1] = [1/1] corresponds to ‘no adjustment’, or ‘multiply frequency by 1/1’
- In mathematical terms: rational commas form a group under multiplication

Here is a summary of how to find notations for free-JI frequencies:

- Start with a rational JI frequency of the form *v*/*w*
- Find the prime factorisation of *v*/*w* which looks like $2^a3^b5^c7^d$… with a finite number of terms. Zero exponents can be ignored.
- Out of these terms, take only the terms for primes 5 and above, these give a set of terms $\{p^k\}$ where $p \geq 5$ and *k* is a non-zero integer.
- The notation for *v*/*w* will include the rational commas in the set $\{[p^k]\}$ which is the same as $\{[p]^k\}$
- Multiply these $[p^k]$ together to get a rational comma of the form [*x*/*y*]
- This [*x*/*y*] corresponds to a frequency shift which is itself a rational number.
- To find the 3-limit component: divide the original frequency *v*/*w* by the frequency shift [*x*/*y*]
- Use modified SPN to notate this 3-limit frequency.
- Combine the modified SPN notation with a rational comma [*x*/*y*] to get a final notation for *v*/*w*

To clarify and help explain this process above, examples will now be given of the forward and reverse processes.

Turning a frequency fraction into a notation: Derive a notation for 20/21. Two higher primes are present, 5 and 7, so the final notation will need to combine two prime commas, a [5] and a [7]. Their powers are 1 and -1 respectively, so the final notation will use the term $[5]^1[7]^{-1}$ which combine to give the rational comma [5/7]. Make two assumptions: assume [5] = 80/81, and assume [7] = 63/64, so [5/7] is (80/81)/(63/64). Dividing the rational comma frequency out of the original frequency:
20/21 / ((80/81)/(63/64)) is with some simplification equal to 243/256. This is the B just below middle C ($C_4$ = 1/1). Hence the Pythagorean 243/256 has notation $B_3$, and notation for 20/21 is obtained from combining $B_3$ with [5/7] to give $B[5/7]_3$.

Turning a notation into a frequency fraction: As an example of the reverse process, derive a frequency for $D[35]_4$. Now Pythagorean $D_4$ is 9/8, and assume again that [5] = 80/81, and [7] = 63/64 which are the factors of the rational comma [35]. Then multiply these all together to get the answer:
(9/8) × (80/81) × (63/64) is 45360/41472, however this simplifies to 35/32. Hence, $D[35]_4$ = 35/32. Also, since $D_4$ = 9/8 = 36/32, both the rational comma [35] and the notation $C[35]_4$ must have frequency 35/36. This is an example of how the notation $C_4$ = 1/1 is the identity element of modified SPN.

The rational commas [*x*/*y*] in this microtonal adjustment scheme have *x*, *y* with no prime factors of 2 or 3. This means that *x*, *y* are ‘5-rough’; they are numbers having only prime factors 5 and above. The 5-rough numbers start 1, 5, 7, 11, 13, 17, 19, 23, 25, 29, 31, 35, 37… (see OEIS 2016, sequence ‘A007310) and correspond to the numbers which are ±1 mod 6.

For 7-limit work (representing much of Western classical and jazz harmony) *x* and *y* are both 5-rough and 7-smooth. This is a limited set of numbers, starting 1, 5, 7, 25, 35, 49, 125… (see OEIS 2016, sequence ‘A003595’). Hence for 7-limit work only a small set of commas are typically needed, such as [5], [25], [1/5], [7], [1/7], [5/7], [7/5], [35]. The 7-limit notation is easier to learn and use, since knowledge of only a limited comma set is required.

## 5) Rational Comma Notation (RCN) to notate the whole of free-JI

In the last two sections, two components of free-JI notation have been described:

- Modified SPN: Notates 3-limit Pythagorean harmony, using notations of the form $L_z$ for L a note label and *z* an octave number
- Rational commas: Notate the set of microtonal adjustments coming from higher primes in a frequency, using notations of the form [*x*/*y*]. Each prime factor *p* of *x* represents a comma component [*p*] in [*x*/*y*], and each prime factor *q* of *y* represents a comma component 1/[*q*] in [*x*/*y*].

Combining these together gives Rational Comma Notation (RCN) which has notations of the form $L[x/y]_z$ and which notates the whole of free-JI.

RCN requires a choice to be made for [5], [7], [11], [13], [17], [19], [23]…, i.e. [*p*] must be chosen for every $p \geq 5$. Since the list of primes is infinite, an algorithm must be specified to assign and manage all of these choices. Denoting the algorithm by X, let this version of RCN be called $RCN_X$.

RCN is a notation style which does not depend on any particular algorithm being chosen. In the following sections, two algorithms are found in the literature (SAG, KG2) and then a novel algorithm is described (DR) which has improved properties in some respects. When using RCN, the choice of algorithm for higher prime commas ought to be stated, e.g. $RCN_{DR}$ or $RCN_{KG2}$, unless a composition uses only prime commas which all known algorithms agree on. Currently [5] and [7] are defined the same in all three algorithms (DR, SAG, KG2) so for 7-limit work it might not be necessary to specify an algorithm.

## 6) JI and free-JI notation systems in the literature

Regarding Just Intonation generally: many accounts have been written regarding the history of musical tuning and harmony; the reader is referred to Fauvel, Flood & Wilson (2006) for a history of tuning and temperament; to Partch (1974) for a history of tuning with Just Intonation in mind; see also Haluska (2004) and Sethares (2005). Other authors with historical overviews include Doty (2002), Downes (2008) and Sabat (2009).

Some existing examples of JI notation, microtonal commas and comma assignment algorithms:

- Helmholtz (1895) gave notation for the overtone scale from 1 to 16.
- Partch (1974) custom-built JI instruments. His notation tended to be customised for that instrument.
- Ben Johnston developed a notation system in the mid-twentieth century, described later by Snyder (2010).
- Sabat (2005) described a system of microtonal accidentals in "The Extended Helmholtz-Ellis JI Pitch Notation" and provided prime commas for primes up to 61.
- Monzo (2016) mentions prime exponent vectors which represent the exponent set of a JI interval as a vector, e.g. $80/81 = 2^4 3^{-4} 5$ as [4, -4 ,1 >

- Secor & Keenan (2012) describe a general system called Sagittal which uses arrows in various formats as accidentals to notate microtonal intervals.
- Kite Giedraitis (2017) uses a colour notation which pairs the names of colours to particular primes and their reciprocals.

Since there are already many JI notation systems, why invent a new one? The main reason is to overcome the limitations of existing systems. The various limitations are categorised below, with definitions as necessary:

- (Not) **Universal**: Let 'universal' be defined as 'having a notation for every prime, no matter how high'. Notation systems not being universal means not having a notation for primes above a certain limit, e.g. 31 or 61. Free-JI notation must be universal.
- (Not) **Algorithmic**: Let 'algorithmic' be defined as 'having an algorithm to assign a notation to every prime'. An algorithm is a finite description about how to assign a prime comma to any given prime. Without an algorithm the algorithm must also not be universal, since any (tractable) list of prime comma assignments is finite, which means some prime commas would be missing. Hence any free-JI notation system must have a clearly defined algorithm.
- **Unwieldy**: Notations which get longer and longer as the prime limit increases, typically by requiring a new entry to represent each new prime. Free-JI notation ought not to be unwieldy.
- (Not) **Plaintext**: Let 'plaintext' be defined as 'being able to write the notation using a computer keyboard', e.g. 'using ASCII characters with codes from 32 to 127'. Notations without a plaintext representation are difficult to enter into a computer sequencer without specialist software, or to discuss in typed format, e.g. email or online discussion forum. Ideally a free-JI notation system should have a plaintext variant available.
- **Translation** problem: Needing to learn new symbols or accidentals for each new prime. Ideally there would never be a translation problem, although any free-JI notation would require users to learn some new concepts. A good free-JI notation would minimise translation problems, although this may be a subjective matter since different individuals may variously prefer numbers, letter, colours, accidentals, arrows or other graphics. One way around this is to have more than one free-JI notation system, with software/online translation available to go between the different systems. This is briefly discussed in section (14) below.
- (Not) **Mixed**: Let 'mixed' be defined as 'notations which define some commas for primes 5 and above, which contain other primes 5 and above'. Conversely, 'not mixed' has every comma of the form $2^a3^bp$; the definitions can be described by the 3-exponent $b$ for each prime, which corresponds to a unique note label for the octave-reduced prime $p/2^k$. An example of a mixed definition: defining the prime comma for 17 as 255/256, since then 17 is mixed with 5. In mixed systems it becomes much more convoluted to map between notations and frequencies. For free-JI, an unmixed system is strongly preferable since it makes mapping between notations and frequencies as easy as possible; examples for forward and reverse mapping were given earlier.
- (Not) **Compact**: Let 'compact' be defined as 'all higher prime information can be summarised by a single notational unit, of equivalent descriptive power as one rational number'. Lack of compactness means that: for each prime 5 and above in a musical interval, there may be no simpler representation of their effects than listing the prime commas separately. This list could theoretically extend indefinitely. However, in a compact notation all higher prime information can always be combined into a single unit which is of comparable length to a rational number. For free-JI, compact is preferable, but not essential.

Here then is an analysis of the existing notation systems listed above, with these properties in mind:

- Helmholtz's notation for overtones from 1 to 16 is not universal. It is not readily deducible how to notate an interval such as 17/16.
- Partch did not provide a general purpose notation at all.
- Johnston's notation system assigned E = 5/4, A = 5/3, B = 15/8; thus the note labels do not follow a series of perfect fifths and the effect of prime 5 is mixed into higher primes for any choice of prime commas. These make the notation system harder to use than an unmixed system.
- The Extended Helmholtz-Ellis notation has an attractive set of accidentals and blends well with traditional stave notation. However, it has several problems in the categories listed above: it is not universal, defining only accidentals up to 31 or 61 (depending on the version); it is not plaintext, hindering online discussion of its notations; it has a translation problem since two new accidental graphics (or two new combinations of accidental graphics) must be learned per prime; it has several mixed commas, two examples include comma 255/256 mixing 17 with 5, and comma 1023/1024 mixing 31 with 11.
- A prime-exponent vector notation (for example [-1,0,0,0,-3,5> = $2^{-1}11^{-3}13^{5}$ in 13-limit JI) is unwieldy. As prime limit increases, new vector components must be added. This notation style is likely to be useful only for smaller prime limits, depending on personal preference for writing out longer vectors; but not useful to notate high prime limits, hindering its usefulness for free-JI.
- Sagittal is a flexible and multipurpose notation system, capable of handling both JI and general microtonal work. It benefits from having a plaintext version available for the set of arrows/accidentals. A recent online discussion thread (Keenan & Taylor 2016) gave an algorithm 'SAG' to universally assign notations using prime comma 3-exponents from -6 to +6 (F# to Gb); algorithm SAG is outlined in section (13) below. The main drawback is the translation problem between higher prime content and a range of arrow head styles, making it hard to know which primes are in a notation without detailed recall of what all the (slightly unusual-looking!) arrow head accidentals mean. Compactness is partial, meaning that in some circumstances the separate accidentals representing higher primes might be combined, although again there is a translation issue of knowing when this might or might not be allowed or advisable, and recognising the components of a combined arrow head.
- Kite's colour notation is based on the Pythagorean series of fifths (white notes) and assigns colours to higher primes: 5 is 'yellow', 1/5 is 'green', 7 is 'blue', 1/7 is 'red'. These can be abbreviated y, g, b, r to use colour names as single-letter accidentals in a plaintext version, e.g. 'wC' for 'white C' which might be assigned to 1/1, 'yE' for 'yellow E' which is then 5/4. Primes 11, 13, 17, 19 and their inverses also receive colours and letters, giving twelve colours and letters to remember. To avoid having too many colours, primes from 23 onwards and their reciprocals are notated similarly to '23i-' '23q-' '29i-' etc. The notation is universal since an algorithm 'KG2' exists to assign 3-exponents from -6 to +6 (F# to Gb) to each prime number, based on the prime's position compared to bands of 50 and 100 cents in the octave; algorithm KG2 is also outlined in section (13) below. (A previous algorithm KG is no longer in use, hence the current algorithm is denoted KG2). This does raise one question about the KG2 algorithm, which is whether the prime would be better compared to twelve consecutive Pythagoreans to classify it (as a retuned Pythagorean) instead of comparing it to the 12-EDO intervals? (The author of this paper had a previous algorithm which compared to 12-EDO, and has now switched to a comparison with twelve consecutive Pythagoreans in the current work, since this obtains 'better' results for some primes such as 11; what is meant by 'better' is discussed in the next section.) The colour notation is not compact, in that each higher prime alteration must be listed as a separate letter or item (e.g. 'yy' for a factor of 25, 'yyy' for a factor of 125), however non-compactness has a fringe benefit of allowing alternate commas and their notations to be defined for some primes, if so desired, benefiting any user who disagrees about the algorithm's default choice for particular primes.

Out of these seven notation systems described, only two had algorithms available (Sagittal, Kite's colour notation, yielding SAG and KG2 algorithms respectively), and so only these two systems will be able to notate free-JI. Moreover, neither SAG or KG2 systems are fully compact in their native notation systems. Their algorithms can however be used with Rational Comma Notation (RCN) to give two versions $RCN_{SAG}$ and $RCN_{KG2}$ which are compact; any version of RCN is compact since two rational commas can always be combined into a single rational comma.

In the next section the criteria for algorithmic selection of prime commas are considered, which leads to a divergence from SAG and KG2 algorithms and to the development of a novel algorithm denoted 'DR'.

## 7) Criteria for selection of prime commas for novel algorithm denoted 'DR'

To recap: under a compact free-JI notation system different prime commas can be always combined into a single rational comma. Also, rational commas can be separated out into their prime components. Thus there must be a unique prime comma for each prime number. Otherwise, there is no way of mapping backwards and forwards between prime numbers and their microtonal adjustments in the JI notation. For example, the rational comma [25/7] requires that [5] and [7] are both defined and both represent a unique microtonal adjustment, otherwise $[25/7] = [5]^2[7]^{-1}$ is ill-defined as a notation.

It follows that the choices for all prime commas must be sensible, and optimal in whichever ways are desirable. Two such ways are:

- [*p*] is microtonal and thus should be approximately equal to 1/1. The closer to 1/1, the better.
- $[p] = 2^a3^bp$ is a rational number with a numerator and a denominator. The smaller these are, the better. Two reasons why:
    - Commas with small numbers are easier to work with, it is easier to translate between notations and frequencies
    - As shown below, it would cause primes near whole-numbered Pythagoreans to receive the same note label as that Pythagorean, e.g. 257 would have the same label as 256. This treats Pythagoreans amongst the whole numbers as the natural source of note labelling.

'Closeness of [$p$] to 1/1' and 'smallness of numerator and denominator in $2^a 3^b p$' generally work against each other; it is often possible to find a larger numerator and denominator which get [$p$] closer to 1/1. There turns out to be a trade-off between these two criteria, and the trade-off can be resolved satisfactorily by quantifying both into suitable metrics and then multiplying them together. Suitable metrics are:

- $AO = |\log_2(2^a 3^b p)|$ 'Absolute Octaves' of how big (in octaves) the interval is
- $LCY = \log_2(2^{|a|} 3^{|b|} p)$ 'Log Complexity' or 'Tenney height' of how big the numbers are which also has units of octaves (see Tenney 1983, Ryan 2016)
- $CM = AO \times LCY$ 'Comma Measure' which is **minimised** for an optimal prime comma (Can derive $CM = |\log_2(n)^2 - \log_2(m)^2|$ where $2^a 3^b p$ in lowest terms is $n/m$)

Now the best choice of $a$ can be calculated from specific values of $p$ and $b$. Since [$p$] is near 1/1, then $2^a 3^b$ must be near $p$. Rearranging, $2^a$ must be near $3^{-b}p$. Hence, the best choice for $a$ is round($\log_2(3^{-b}p)$) since $a$ is a whole number.

It follows that the unique prime comma [$p$] must be chosen from a one-dimensional list of candidate commas $2^a 3^b p$ where the list is enumerated using integer values of $b$ and then each value of $a$ calculated from $p$ and $b$. For each prime, its list of candidate commas will have different values for $CM$, and the candidate with the smallest $CM$ can then be defined as the optimal prime comma for that prime number.

For the candidate commas, $b$ could potentially take any integer value ($a$ will take not all integers but a fraction $(\log_2 3)^{-1} = 0.6309\ldots$ of them). But should all integer values of $b$ be used? No, since larger values of $a$, $b$ are 'unmusical'. An example is:

- Suppose the comma [5] was assigned as 80/81, then $E[5]_4 = 5/4$
- Suppose instead that [5] = 32805/32768, then $Fb[5]_4 = 5/4$
- These follow from a Pythagorean $E_4$ being 81/64, and Pythagorean $Fb_4$ being 8192/6561, as set out in Table 1 above.

But all musicians know that a major third above C is labelled 'E' and not 'Fb'. Hence in Just Intonation if 1/1 is a type of C note, then 5/4 must be a type of E note, and cannot simultaneously be a type of Fb note, since E and Fb are distinct in modified SPN. So even though the candidate comma 80/81 has a larger (less optimal) value of $CM$ than 32805/32768, even so 80/81 should be preferred due to musical considerations – prime commas are being chosen for musical use, not for mathematical interest.

Hence $b$ will not range over the whole of the integers, but only be taken from a finite set of integers of musical relevance. Due to there being no apparent reason why this set should have any gaps, the simplest solution is for $b$ to range over a set of consecutive integers. This range was fixed (-6 to +6) in the two algorithms (SAG, KG2) from the literature, however an argument will be made below for a dynamic range which varies according to the prime $p$, and tending towards negative values of $b$, since positive values give larger comma numerators and denominators.

The range can be constructed if two questions are answered: how many consecutive integers should be chosen for the range of $b$, and what is the midpoint of this range?

Suppose $b$ is chosen from $N$ consecutive integer values. The candidate commas form relative pitch classes equivalent to the set $\{3^b: b = b_{min}\ldots b_{max} = 0\ldots N\text{-}1\}$, which can be mapped to pitch classes in the octave [1/1, 2/1), reordered into ascending order, and giving a scale of $N$ Pythagorean pitch classes in the octave. As an example, {1, 3, 9, 27} map into the octave as {1/1, 3/2, 9/8, 27/16}, and are reordered in ascending order as (1/1, 9/8, 3/2, 27/16, 2/1) where 2/1 has been added to complete the scale and give $N$ gaps. So what properties does this type of Pythagorean scale generally have? The key property in this application is the gap structure between consecutive scale notes.

Each prime number also gives rise to a pitch class in this scale (if $b_{min} = 0$) or a transposed version of it (if $b_{min}$ varies) and a desirable property of the scale would be that each scale note approximates roughly the same number of primes, i.e. that the prime pitch classes fall evenly onto scale pitch classes. This only happens if the gaps in the scale are all approximately the same size. It turns out that for a special series of numbers the gap size variation is much lower than for other numbers. That special series is the list of denominators of convergents for the continued fraction series of $\log_2 3 = 1.58496\ldots$ which starts 1, 2, 5, 12, 41, 53, 306, 665, 15601, 31876… (see OEIS 2016, 'sequence A005664').

So – which value of scale length $N$ to choose? The answer is, of course, to choose 12! But the reasoning is subtle, and not dependent on 12-EDO being the current cultural tuning of choice. Here is the reasoning point by point:

- Choosing $N = 12$ gives a very even Pythagorean scale with 5 gaps of size 113.69 cents and 7 gaps of size 90.22 cents. If the central values of $b$ gave C and G then the whole scale would contain Db, Ab, Eb, Bb, F, C, G, D, A, E, B, F# Pythagorean pitch classes. For each prime, there would be a candidate comma of size less than 57 cents, which is within the microtonal range. So $N = 12$ is acceptable. (Note that this candidate comma is not necessarily optimal – however as seen later the (probable) largest optimal comma under the chosen algorithm is [13] = 26/27 at 65.34 cents.)
- Choosing $N = 11$ gives a scale with one gap of 204 cents. This means some prime numbers could be up to 102 cents from a candidate comma. Since [$p$] must be microtonal, potential scale gaps of over a semitone are unacceptable. so $N = 11$ should be rejected in favour of $N = 12$.
- Since for any $N < 12$ there will be a gap at least 204 cents, any $N < 12$ should be rejected in favour of $N = 12$.
- If $N$ was restricted to only $\log_2 3$ denominators of convergents to create an evenly spread scale, the next choice is $N = 41$. However, this means that the Pythagorean scale centred on C will run over 41 labels from Gbbb to F###. These labels use far more sharps and flats than $N = 12$, and such labels are unmusical, leading to effects such as 5/4 being labelled an Fb. It is also much less convenient for the musician to have to deal with 41 labels rather than 12. So mainly by considerations of musicality, but also of usability and convenience, reject $N = 41$ in favour of $N = 12$.
- Since any other higher $\log_2 3$ denominators of convergents (53, 306, 665, 15601…) will also give an unmusical (and unusable/inconvenient) number of sharps and flats, reject all such values in favour of $N = 12$.
- Consider smaller numbers which are not $\log_2 3$ denominators of convergents, in the range 13…40. Suppose $N = 13$. Then there would be a gap of 23 cents in the scale. For example, the 13-note scale from Gb to F# has Gb and F# at 23 cents apart. Moreover, Gb is lower than F#. Hence, the labels get out of sequence (see Table 2 above), so part of the scale would read 'F, Gb, F#, G'. This undesirable sequencing, combined with the uneven gap sizes of the scale, make $N = 12$ a better choice than $N = 13$.

- Similar reasoning will reject any other $N = 13 \ldots 40$ in favour of $N = 12$.
- The same is true for any higher $N > 41$; reject it for $N = 12$.
- Hence by a process of elimination, if a fixed length of Pythagorean scale is being used to approximate pitch classes from primes, use the scale length $N = 12$.

At this point the design choices for algorithm DR have diverged from both SAG and KG2, which both have $N = 13$; for SAG this means that some note labels are out of sequence and the note labels are not used evenly; for KG2 some note labels would be out of sequence were it not for a reordering within the algorithm, which has the effect of having some ranges within the octave represented by a Pythagorean fraction outside of those ranges (e.g. having 550 to 600 cents represented by an F#, 729/512 at 611.73 cents).

So for the novel algorithm DR, $N = 12$, and $b$ takes by default 12 values. What should the midpoint $b_{\text{mid}}$ of these values be? Since $[p] = 2^a3^bp$ is approximately equal to 1, at the midpoint the contributions of $a$ and $b$ in $[p]$ should be approximately the same size. This leads to $2^a \approx 3^b \approx p^{-2}$, and by taking logarithms base-3 of the right hand approximation, the midpoint is found to be $b_{\text{mid}} = -\log_3(p)/2$. Since this number is irrational, take the 12 integer values of $b$ which are closest to $b_{\text{mid}}$, which are $|b - b_{\text{mid}}| < 6$.

Are 12 values of $b$ enough to find suitable candidate commas for all prime numbers? For small prime numbers, yes. For larger prime numbers, no. Consider a large prime number $p$ which was of the form $2^x \pm k$, $3^y \pm k$ or $2^x3^y \pm k$ where $k$ was very small in comparison to $2^x$, $3^y$ or $2^x3^y$ respectively, and these $x$, $y$ positive integers. In such cases, the best candidate comma for the prime number ought to be of the form $2^{-x}p$, $3^{-y}p$ or $2^{-x}3^{-y}p$ respectively. This leads to extending the 12 candidate commas slightly for large prime numbers so that any $2^a3^bp$ where $p$ is the numerator becomes a candidate comma. It turns out that up to prime 375,779 only the 12 candidate commas described previously are needed, but for primes from $P_{\text{B}} = 375{,}787$ there are 13 or more candidate commas with $p$ as the numerator.

Some terminology for these: the 'primary range' (PR) being the candidate commas with $p$ as the numerator, and the 'secondary range' (SR) being the 12 nearest candidates to the midpoint $b_{\text{mid}}$. Two facts which can be shown are: one of these ranges (PR, SR) is always a subset of the other, so the two ranges agree; and for large primes, the number of candidate commas in PR grows at approximately $\log_3(p)$.The secondary range SR is only needed for smaller primes than $P_{\text{B}}$, the primary range PR is used for all subsequent primes. The algorithm can thus be summarised as:

**'Minimise *CM* over: SR for primes below $P_{\text{B}}$, or over PR for primes from $P_{\text{B}}$ onwards'**

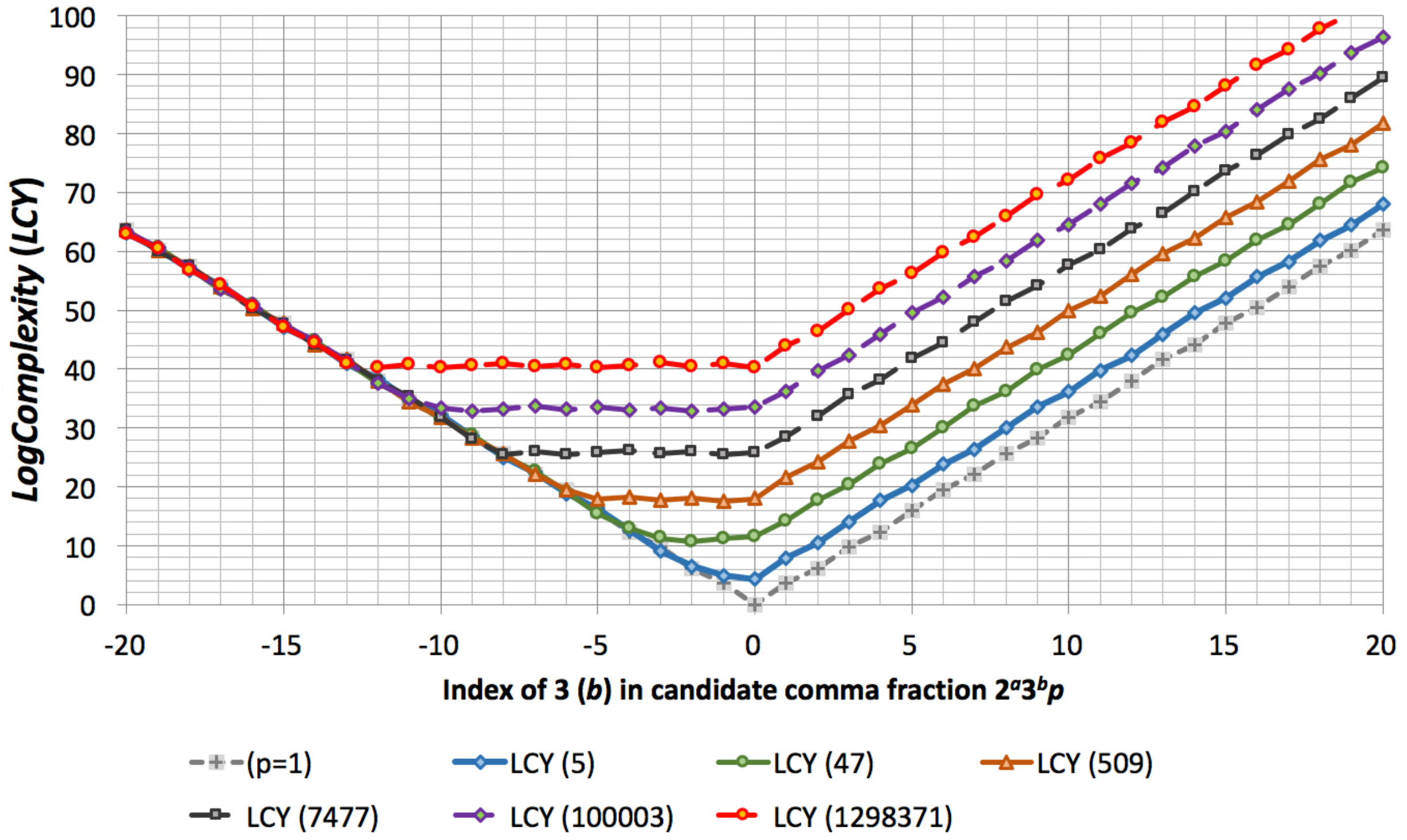

**Figure 1: Graph of *LogComplexity* (*LCY*) of candidate commas vs Index of 3 (*b*) for candidate commas of the form $2^a3^bp$, for six primes between 5 and 1.3 million, also with a dummy data series with *p* set to 1**

Figure 1 gives a useful graphical representation of the candidate commas $2^a3^bp$ for varying $p$ and $b$. Six series have been plotted for six primes, roughly evenly spaced on a log-scale from 5 to 1.3 million. A seventh (dummy) series has been plotted for $p = 1$, which shows symmetry on left and right and is centered on (0, 0). On the horizontal axis, $b$ has been varied between -20 and +20. For each ($p$, $b$) pair, a candidate comma has been calculated, and its *LogComplexity* (*LCY*) has been plotted on the vertical axis. These give the seven data series above.

Immediately the 'shape' of each series of candidate commas for each prime becomes apparent; each series splits into three regions. These can be denoted 'left', 'middle' and 'right' in relation to the graph above. They represent three distinct types of candidate comma: $2^{|a|}p/3^{|b|}$, $p/(2^{|a|}3^{|b|})$, $3^{|b|}p/2^{|a|}$ respectively. The middle region with candidates $p/(2^{|a|}3^{|b|})$ has an approximately flat *LCY* curve, where *LCY* is minimal and lower than in the left or right regions.

Here are some properties which hold in the graph's middle region for each prime $p$:

- It has all of the candidate commas with $p$ in the numerator
- The 'middle region' is therefore the same range as PR
- Numerator and denominator are minimised in PR
- *LCY* is minimised in PR
- PR contains good candidate commas, since low *LCY* value was listed earlier as one of the two desirable features of a prime comma
- The centre of PR is very close (up to a small constant factor) to the centre of SR which is $b_{\text{mid}}$
- There are approximately the same number of candidate commas in PR to the left and right of $b_{\text{mid}}$

The conclusion is that all of the candidate commas in the middle region (PR) ought to be searched in the algorithm, as well as twelve candidates either side of $b_{mid}$ (SR). Since PR varies according to $p$, a fixed search range (such as -6 to +6) will not be optimal for all primes. Since existing algorithms SAG, KG2 only use fixed search ranges, this supports development of a novel algorithm DR with a dynamic search range.

Overall then, clear guidelines have been described for how candidate commas ought to be selected for each prime number, and how an optimal prime comma ought to be selected from these candidates. These guide the construction of algorithm DR, presented below.

## 8) Presentation of novel algorithm (DR) to determine prime commas

In this section the algorithm DR is given for first discovering the set of candidate commas for a prime number, and then for selecting the optimal prime comma from the candidates which minimises the *CM* value. Here is the one sentence summary of the algorithm, repeated from earlier:

**'Minimise *CM* over: SR for primes below $P_B$, or over PR for primes from $P_B$ onwards'**

In this sentence, *CM* is the *CommaMeasure* function, SR is the secondary range of 12 candidate commas closest to $b_{mid}$, PR is the primary range of candidate commas with $p$ in the numerator, $P_B$ = 375,787 is the boundary where PR overtakes SR. Here is the algorithm in full, which unpacks the definitions above:

Choose a prime number $p$ (its candidate commas are all of the form $1/1 \approx 2^a 3^b p$ for integer $b$)

Calculate: $b_{mid} = -\log_3(p)/2$ (unlike values $b_x$ below, this $b_{mid}$ is not an integer)

Calculate: $b_{min1} = \text{round}(b_{mid} - (12-1)/2)$

Calculate: $b_{max1} = \text{round}(b_{mid} + (12-1)/2)$ (these give SR)

Calculate: $b_{min2} = \text{ceiling}(-\log_3(p) - 1/(2\log_2 3))$

Calculate: $b_{max2} = 0$ (these give PR)

Calculate: $b_{min} = \min(b_{min1}, b_{min2})$

Calculate: $b_{max} = \max(b_{max1}, b_{max2})$ (final range is union of the previous two ranges)

**Loop** $b = b_{min}$ to $b_{max}$

Calculate: $a = \text{round}(-\log_2(p) - b\log_2 3)$ (value of $a$ to make candidate closest to 1/1)

Calculate: $[p]_{cand} = 2^a 3^b p$ (microtonal frequency of candidate interval)

Calculate: $AO = |\log_2(2^a 3^b p)|$ (Absolute Octave size of candidate interval)

Calculate: $LCY = \log_2(2^{|a|} 3^{|b|} p)$ (Log Complexity in octaves / Tenney Height)

Calculate: $CM = AO \times LCY$ (Comma Measure for optimisation)

Save $p$, $a$, $b$, $[p]_{cand}$, $AO$, $LCY$, $CM$ into a suitable table

**End loop**

A table has been obtained in which each candidate comma gives a new row

Sort the table by the column *CM* in increasing order

The candidate comma $[p]_{\text{cand}}$ with minimal *CM* is now in the first row of the table

**Define [*p*] to be this $[p]_{\text{cand}}$** found on the first row – so the unique and optimal prime comma for *p* is the candidate comma found on the first row of this calculation table which has minimal *CM* across all the selected candidates.

Repeat the algorithm above for different prime numbers, as required

## 9) Results for novel algorithm (DR) – prime commas for primes below 1400

Algorithm DR from the section above was implemented and initially run for prime numbers up to 200. Values for *p*, *a*, *b*, $[p]_{\text{cand}}$, *AO*, *LCY* and *CM* were recorded for all candidate commas, and for each prime number these candidate commas were sorted by ascending *CM*. The candidate comma with minimal *CM* was retained for each prime number, and presented in Table 5 below. Note that all of these columns including the 3-exponent *b* are independent of the convention $C_4 = 1/1$, except for the final column 'Pitch Class label of *p*' showing how the pitch class of each prime is a Pythagorean pitch class with [*p*] added:

**Table 5: Prime commas $[p] = 2^a3^bp$ and supporting information, for DR algorithm and *p* below 200**

| *p* | [*p*] fraction | [*p*] cents | [*p*] decimal | *LCY* | *AO* | *CM* | *a* | *b* | Pitch Class label of *p* |
|---|---|---|---|---|---|---|---|---|---|
| 5 | 80/81 | -21.51 | 0.9877 | 12.662 | 0.018 | 0.227 | 4 | -4 | E[5] |
| 7 | 63/64 | -27.26 | 0.9844 | 11.977 | 0.023 | 0.272 | -6 | 2 | Bb[7] |
| 11 | 33/32 | 53.27 | 1.0313 | 10.044 | 0.044 | 0.446 | -5 | 1 | F[11] |
| 13 | 26/27 | -65.34 | 0.9630 | 9.455 | 0.054 | 0.515 | 1 | -3 | A[13] |
| 17 | 2176/2187 | -8.73 | 0.9950 | 22.182 | 0.007 | 0.161 | 7 | -7 | C#[17] |
| 19 | 513/512 | 3.38 | 1.0020 | 18.003 | 0.003 | 0.051 | -9 | 3 | Eb[19] |
| 23 | 736/729 | 16.54 | 1.0096 | 19.033 | 0.014 | 0.262 | 5 | -6 | F#[23] |
| 29 | 261/256 | 33.49 | 1.0195 | 16.028 | 0.028 | 0.447 | -8 | 2 | Bb[29] |
| 31 | 31/32 | -54.96 | 0.9688 | 9.954 | 0.046 | 0.456 | -5 | 0 | C[31] |
| 37 | 37/36 | 47.43 | 1.0278 | 10.379 | 0.040 | 0.410 | -2 | -2 | D[37] |
| 41 | 82/81 | 21.24 | 1.0123 | 12.697 | 0.018 | 0.225 | 1 | -4 | E[41] |
| 43 | 129/128 | 13.47 | 1.0078 | 14.011 | 0.011 | 0.157 | -7 | 1 | F[43] |
| 47 | 47/48 | -36.45 | 0.9792 | 11.140 | 0.030 | 0.338 | -4 | -1 | G[47] |
| 53 | 53/54 | -32.36 | 0.9815 | 11.483 | 0.027 | 0.310 | -1 | -3 | A[53] |
| 59 | 236/243 | -50.60 | 0.9712 | 15.807 | 0.042 | 0.667 | 2 | -5 | B[59] |
| 61 | 244/243 | 7.11 | 1.0041 | 15.856 | 0.006 | 0.094 | 2 | -5 | B[61] |
| 67 | 2144/2187 | -34.38 | 0.9803 | 22.161 | 0.029 | 0.635 | 5 | -7 | C#[67] |
| 71 | 71/72 | -24.21 | 0.9861 | 12.320 | 0.020 | 0.249 | -3 | -2 | D[71] |
| 73 | 73/72 | 23.88 | 1.0139 | 12.360 | 0.020 | 0.246 | -3 | -2 | D[73] |
| 79 | 79/81 | -43.28 | 0.9753 | 12.644 | 0.036 | 0.456 | 0 | -4 | E[79] |
| 83 | 83/81 | 42.23 | 1.0247 | 12.715 | 0.035 | 0.447 | 0 | -4 | E[83] |
| 89 | 712/729 | -40.85 | 0.9767 | 18.986 | 0.034 | 0.646 | 3 | -6 | F#[89] |
| 97 | 97/96 | 17.94 | 1.0104 | 13.185 | 0.015 | 0.197 | -5 | -1 | G[97] |
| 101 | 6464/6561 | -25.79 | 0.9852 | 25.338 | 0.021 | 0.544 | 6 | -8 | G#[101] |
| 103 | 6592/6561 | 8.16 | 1.0047 | 25.366 | 0.007 | 0.173 | 6 | -8 | G#[103] |
| 107 | 107/108 | -16.10 | 0.9907 | 13.496 | 0.013 | 0.181 | -2 | -3 | A[107] |
| 109 | 109/108 | 15.96 | 1.0093 | 13.523 | 0.013 | 0.180 | -2 | -3 | A[109] |
| 113 | 1017/1024 | -11.88 | 0.9932 | 19.990 | 0.010 | 0.198 | -10 | 2 | Bb[113] |
| 127 | 127/128 | -13.58 | 0.9922 | 13.989 | 0.011 | 0.158 | -7 | 0 | C[127] |
| 131 | 131/128 | 40.11 | 1.0234 | 14.033 | 0.033 | 0.469 | -7 | 0 | C[131] |
| 137 | 2192/2187 | 3.95 | 1.0023 | 22.193 | 0.003 | 0.073 | 4 | -7 | C#[137] |
| 139 | 2224/2187 | 29.04 | 1.0169 | 22.214 | 0.024 | 0.538 | 4 | -7 | C#[139] |
| 149 | 4023/4096 | -31.13 | 0.9822 | 23.974 | 0.026 | 0.622 | -12 | 3 | Eb[149] |
| 151 | 4077/4096 | -8.05 | 0.9954 | 23.993 | 0.007 | 0.161 | -12 | 3 | Eb[151] |
| 157 | 157/162 | -54.28 | 0.9691 | 14.634 | 0.045 | 0.662 | -1 | -4 | E[157] |
| 163 | 163/162 | 10.65 | 1.0062 | 14.689 | 0.009 | 0.130 | -1 | -4 | E[163] |
| 167 | 501/512 | -37.60 | 0.9785 | 17.969 | 0.031 | 0.563 | -9 | 1 | F[167] |
| 173 | 519/512 | 23.51 | 1.0137 | 18.020 | 0.020 | 0.353 | -9 | 1 | F[173] |
| 179 | 716/729 | -31.15 | 0.9822 | 18.994 | 0.026 | 0.493 | 2 | -6 | F#[179] |
| 181 | 724/729 | -11.91 | 0.9931 | 19.010 | 0.010 | 0.189 | 2 | -6 | F#[181] |
| 191 | 191/192 | -9.04 | 0.9948 | 15.162 | 0.008 | 0.114 | -6 | -1 | G[191] |
| 193 | 193/192 | 8.99 | 1.0052 | 15.177 | 0.007 | 0.114 | -6 | -1 | G[193] |
| 197 | 197/192 | 44.51 | 1.0260 | 15.207 | 0.037 | 0.564 | -6 | -1 | G[197] |
| 199 | 199/192 | 61.99 | 1.0365 | 15.222 | 0.052 | 0.786 | -6 | -1 | G[199] |

The same algorithm DR was then run for primes $p$ up to 1400. The optimal values of $b$ were noted and presented in Table 6 below, with separate pairs of ($p$, $b$) distinguished by column shading. This table functions as a quick lookup table; given a prime $p$, look up its value for $b$; then all these facts about the comma can be calculated: pitch class label, 2-exponent $a$, prime comma fraction, comma statistics. (A graphical plot of these $b$ values is included in Figure 2 below.)

One feature of Table 6 is that consecutive primes usually have the same $b$ value; this is because the same Pythagorean would usually be the optimal approximation for primes which are only a few cents apart, and as the primes get larger then consecutive primes become closer in terms of cents.

**Table 6: Values of 3-exponent *b* for prime commas [*p*] = $2^a3^bp$; for DR algorithm and *p* below 1400**

| *p* | *b* | *p* | *b* | *p* | *b* | *p* | *b* | *p* | *b* | *p* | *b* | *p* | *b* | *p* | *b* | *p* | *b* | *p* | *b* |
|---|---|---|---|---|---|---|---|---|---|---|---|---|---|---|---|---|---|---|---|
| 5 | -4 | 97 | -1 | 211 | -3 | 347 | 1 | 467 | 2 | 617 | 3 | 761 | -1 | 919 | 2 | 1063 | -7 | 1229 | -9 |
| 7 | 2 | 101 | -8 | 223 | 2 | 349 | 1 | 479 | -5 | 619 | 3 | 769 | -1 | 929 | 2 | 1069 | -7 | 1231 | -9 |
| 11 | 1 | 103 | -8 | 227 | 2 | 353 | -6 | 487 | -5 | 631 | -4 | 773 | -1 | 937 | -5 | 1087 | -7 | 1237 | -9 |
| 13 | -3 | 107 | -3 | 229 | 2 | 359 | -6 | 491 | -5 | 641 | -4 | 787 | -1 | 941 | -5 | 1091 | -7 | 1249 | -9 |
| 17 | -7 | 109 | -3 | 233 | 2 | 367 | -6 | 499 | 0 | 643 | -4 | 797 | -1 | 947 | -5 | 1093 | -7 | 1259 | -4 |
| 19 | 3 | 113 | 2 | 239 | -5 | 373 | -6 | 503 | 0 | 647 | -4 | 809 | -8 | 953 | -5 | 1097 | -7 | 1277 | -4 |
| 23 | -6 | 127 | 0 | 241 | -5 | 379 | -1 | 509 | 0 | 653 | -4 | 811 | -8 | 967 | -5 | 1103 | -7 | 1279 | -4 |
| 29 | 2 | 131 | 0 | 251 | 0 | 383 | -1 | 521 | 0 | 659 | -4 | 821 | -8 | 971 | -5 | 1109 | -7 | 1283 | -4 |
| 31 | 0 | 137 | -7 | 257 | 0 | 389 | -1 | 523 | 0 | 661 | -4 | 823 | -8 | 977 | -5 | 1117 | -7 | 1289 | -4 |
| 37 | -2 | 139 | -7 | 263 | 0 | 397 | -1 | 541 | -7 | 673 | 1 | 827 | -8 | 983 | -5 | 1123 | -2 | 1291 | -4 |
| 41 | -4 | 149 | 3 | 269 | -7 | 401 | -8 | 547 | -7 | 677 | 1 | 829 | -8 | 991 | -5 | 1129 | -2 | 1297 | -4 |
| 43 | 1 | 151 | 3 | 271 | -7 | 409 | -8 | 557 | -7 | 683 | 1 | 839 | -3 | 997 | -5 | 1151 | -2 | 1301 | -4 |
| 47 | -1 | 157 | -4 | 277 | -7 | 419 | -3 | 563 | -2 | 691 | 1 | 853 | -3 | 1009 | 0 | 1153 | -2 | 1303 | -4 |
| 53 | -3 | 163 | -4 | 281 | -2 | 421 | -3 | 569 | -2 | 701 | 1 | 857 | -3 | 1013 | 0 | 1163 | -2 | 1307 | -4 |
| 59 | -5 | 167 | 1 | 283 | -2 | 431 | -3 | 571 | -2 | 709 | -6 | 859 | -3 | 1019 | 0 | 1171 | -2 | 1319 | -4 |
| 61 | -5 | 173 | 1 | 293 | -2 | 433 | -3 | 577 | -2 | 719 | -6 | 863 | -3 | 1021 | 0 | 1181 | -2 | 1321 | -4 |
| 67 | -7 | 179 | -6 | 307 | 3 | 439 | -3 | 587 | -2 | 727 | -6 | 877 | -3 | 1031 | 0 | 1187 | -2 | 1327 | -4 |
| 71 | -2 | 181 | -6 | 311 | 3 | 443 | -3 | 593 | -2 | 733 | -6 | 881 | -3 | 1033 | 0 | 1193 | -2 | 1361 | 1 |
| 73 | -2 | 191 | -1 | 313 | -4 | 449 | 2 | 599 | 3 | 739 | -6 | 883 | -3 | 1039 | 0 | 1201 | -9 | 1367 | 1 |
| 79 | -4 | 193 | -1 | 317 | -4 | 457 | 2 | 601 | 3 | 743 | -6 | 887 | -3 | 1049 | 0 | 1213 | -9 | 1373 | 1 |
| 83 | -4 | 197 | -1 | 331 | -4 | 461 | 2 | 607 | 3 | 751 | -1 | 907 | 2 | 1051 | 0 | 1217 | -9 | 1381 | 1 |
| 89 | -6 | 199 | -1 | 337 | 1 | 463 | 2 | 613 | 3 | 757 | -1 | 911 | 2 | 1061 | -7 | 1223 | -9 | 1399 | 1 |

## 10) Some observations on these optimal prime commas

For each prime number *p*, the main output of algorithm DR is the 3-exponent *b*, representing the match between a prime number and a Pythagorean number. Everything else of interest, such as the comma fraction, the statistics, and the note label for the prime, can be calculated once *b* is known.

Any new implementation of an algorithm can be checked by outputting a table of (*p*, *b*) values (similar to Table 6 above) and checking that these values agree with previous sources, e.g. that for each *p* in both data sources the value of *b* is the same. The author has used this method to verify two independent algorithm DR implementations (in Excel and Matlab/Octave) for primes up to 16000, giving confidence that both have been implemented correctly and that the results given above are reliable.

**Table 7: First prime number $p_{min}$ to receive each 3-exponent $b$ and each note label; $p_{max}$ supplied if it exists.**

| $b$ | +3 | +2 | +1 | 0 | -1 | -2 | -3 | -4 | -5 |
|---|---|---|---|---|---|---|---|---|---|
| **Label** | Eb | Bb | F | C | G | D | A | E | B |
| $p_{min}$ | 19 | 7 | 11 | 31 (or 1, 2) | 47 (or 3) | 37 | 13 | 5 | 59 |
| $p_{max}$ | 619 | 3739 | 45,077 | NA | NA | NA | NA | NA | NA |
| $b$ | -6 | -7 | -8 | -9 | -10 | -11 | -12 | -13 | -14 |
| **Label** | F# | C# Dbp | G# Abp | D# Ebp | A# Bbp | E# Fp | B# Cp | F## Gp | C## Dp |
| $p_{min}$ | 23 | 17 | 101 | 1201 | 7177 | 85,817 | 527,869 | 1,583,591 | 4,750,679 |

In Table 7 the lowest prime numbers $p_{min}$ are given which receive each value of $b$ and its note label. The highest possible value for $b_{max}$ is +5, at prime 5. This would correspond to a candidate comma with note label Db. However, due to the positions of the octave-reduced primes within the octave, the highest actual $b$ value achieved for any prime comma is +3. This corresponds to an Eb, and occurs at primes 19, 149, 151, 307, 311, 599, 601, 607, 613, 617 and 619 only. For $b$ = +3, +2, +1 the note labels are Eb, Bb, F respectively and these occur a finite number of times each. For $b$ = 0, -1, -2, -3… the note labels are C, G, D, A… and these all occur an infinite number of times.

One explanation for this is otonality. The (otonal) harmonic series is 1/1, 2/1, 3/1, 4/1… which contains both integer Pythagoreans of positive offset (3/1, 9/1, 27/1…) and prime numbers (5/1, 7/1, 11/1…). Since primes are otonal, matching them with otonal Pythagoreans makes more sense in general, indeed algorithm DR does this through considering all of the candidate commas in the primary range PR. With the convention of 1/1 being a C, the set of otonal labels starts C, G, D, A, E… and these will be the labels generally used for primes. The finite set of exceptions where primes receive utonal labels (F, Bb, Eb) are due to SR extending from the otonal midpoint $b_{mid}$ into 'utonal territory' due to requiring a minimum of 12 candidate commas for each prime.

In Table 7 from C# onwards some alternative note labels have been given using 'p', the shorthand for a Pythagorean comma 531441/524288, such as B# = Cp, C## = Dp, etc. The table gives the first E#, B#, F## and C##; it may however be desirable to use notations Fp, Cp, Gp, Dp instead for these which eliminate double sharps and the unusual labels E# and B#. The table shows that double sharps (F## onwards) only occur for primes greater than 1.5 million, which means that for most primes likely to be used in musical composition the decision about how to represent double sharps need not be considered.

**Table 8: Largest prime commas found for $p$ < 100,000**

| $p$ | 13 | 797 | 937 | 2389 | 199 | 7159 | 1877 | 1193 | 313 |
|---|---|---|---|---|---|---|---|---|---|
| **[$p$]** | 26/27 | 797/768 | 937/972 | 2389/2304 | 199/192 | 7159/6912 | 1877/1944 | 1193/1152 | 313/324 |
| **Size (cents)** | -65.34 | +64.17 | -63.49 | +62.72 | +61.99 | +60.79 | -60.72 | +60.54 | -59.80 |

Prime commas were searched for $p$ below 100,000 and sorted by comma size $AO$; the nine largest commas found are listed in Table 8. The largest prime comma found was [13] = 26/27, at -65.34 cents, or approximately 2/3 of a semitone. It is likely to be the largest comma of all; above $P_B$ the Pythagorean scales start getting larger than 12 notes, at that point the gaps get smaller. Moreover, for larger primes $p$ is always in the numerator and $LCY$ is approximately constant, so $CM$ is determined mainly by $AO$, and the smallest comma (which is within 57 cents of the prime) would tend to be optimal. Hence it is not expected that a larger optimal comma than [13] = 26/27 will be found for any other prime.

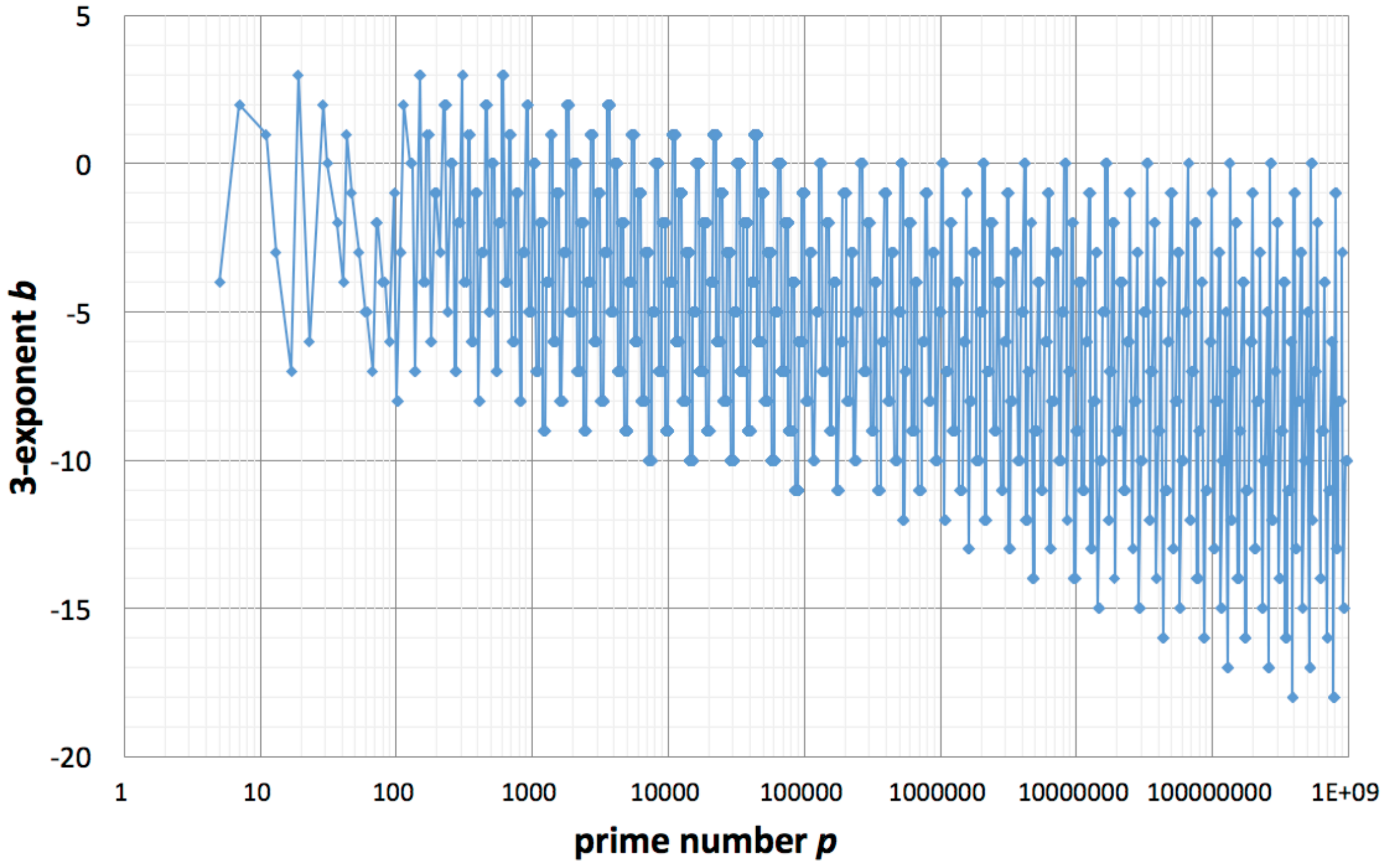

**Figure 2: Graph of 3-exponent (*b*) of comma vs prime (*p*)**

In Figure 2 a graph of 3-exponent $b$ is given for a large range of $p$ values below $10^9$. Minimum and maximum values of $b$ can be read off for each range of $p$; assisted by Table 7 above, $p$ = 45,077 is the last prime with a positive value of $b$. For very small primes $b$ appears random, but a regular order appears for primes above approximately 500. Subsequently $b$ cycles between 12 values as $p$ increases, but above $P_B$ = 375,787 then $b$ cycles between more than 12 (non-positive) values and takes a fixed maximum value 0. As $p$ increases to $10^9$, $b$ is seen to cycle between more and more values.

The 3-exponents of the Pythagorean and the prime comma are negatives of each other. This means that given $b$ for the prime comma, the Pythagorean note name has 3-exponent -$b$. Some examples from Table 5:

- 191, 193, 197 all have $b$ = -1 and are all near 192. Now 192/128 = 3/2 which is a (Pythagorean) G and which has 3-exponent of 1. So b = -1 implies 191, 193, 197 are all microtonal adjustments of a G.

- 127 and 131 ($b = 0$) are both microtonal adjustments of a C since they are near 128, a power of 2 which has 3-exponent of 0.
- 71 and 73 ($b = -2$) are both near 72; 72/64 = 9/8 which is a D with 3-exponent of 2. So 71 and 73 are both microtonal adjustments of a D.

Prime commas [*p*] are always much less than one semitone, however rational commas may be more than one semitone. For example, [13] is -65 cents, and [31] is -55 cents. Together, [13] × [31] = [403] is -120 cents. Additionally, $[169] = [13^2] = [13]^2 = (26/27)^2 = 676/729 = -130.67$ cents. However, this is unusual, since most prime commas being combined are either smaller than these examples, or their directions cancel out. So most rational commas will be less than a semitone, especially for the type of rational comma of greatest practical use, which have smaller numbers and only a few prime factors.

So all prime commas are microtonal, and most (but not all) useful rational commas are microtonal. The important thing is that the rational commas are (almost always) small in relation to the Pythagorean component, which is the case under this scheme for Pythagorean notes not too near $C_4 = 1/1$.

It is possible for *x*/*y* to equal [*x*/*y*]. Consider [73] = 73/72 and [71] = 71/72. Then [73/71] = (73/72) / (71/72) and hence [73/71] = 73/71. Also, [61/59] = 61/59 since [61] = 244/243 and [59] = 236/243. In fact, rational commas for which *x*/*y* equals [*x*/*y*] are closed under multiplication, hence form a subgroup of the whole set of rational commas. An interesting item of further work would be to ascertain exactly which commas these were, and to find which ones have the lowest prime limits. It is hypothesised that there are no such commas in 5-limit or 7-limit, but that one of the simplest in 11-limit is $[12005/14641] = [5][7]^4[11]^{-4} = (80/81)(63/64)^4(33/32)^{-4} = 12005/14641$.

**Table 9: Statistics *3EPO* and *CSPO* for low primes *p***

| ***p*** | 5 | 7 | 11 | 13 | 17 | 19 | 23 | 29 | 31 | 37 |
|---|---|---|---|---|---|---|---|---|---|---|
| ***b*** | -4 | 2 | 1 | -3 | -7 | 3 | -6 | 2 | 0 | -2 |
| ***3EPO*** | **-1.723** | **0.712** | 0.289 | -0.811 | -1.713 | 0.706 | -1.326 | 0.412 | 0.000 | -0.384 |
| **[*p*] in cents** | -21.51 | -27.26 | 53.27 | -65.34 | -8.73 | 3.38 | 16.54 | 33.49 | -54.96 | 47.43 |
| ***CSPO*** | -9.262 | -9.712 | **15.399** | **-17.657** | -2.136 | 0.795 | 3.657 | 6.893 | -11.095 | 9.105 |

In Table 9 two extra statistics are considered, *3EPO* and *CSPO*:

- $3EPO = b / \log_2(p)$ ‘3-Exponent Per Octave’
- $CSPO = 1200 \times \log_2([p]) / \log_2(p)$ ‘Comma Size (in cents) Per Octave’

*3EPO* and *CSPO* are measures of how quickly 3-exponent *b* and comma pitch shift [*p*] in cents deviate from 0, when moving through the integer prime powers $1, p, p^2, p^3 \ldots$. Scaling by $\log_2(p)$ enables a meaningful measure between different primes. Hence *3EPO* and *CSPO* compare the relative speeds by which different primes cause 3-exponent and comma size to deviate.

The minimum and maximum values for these two statistics were taken across all primes between 5 and 100,000, and also a selection of 51,110 primes below 1 billion. For large primes *3EPO* and *CSPO* tended to have small absolute values, which can be explained by both *b* and [*p*] being bounded, but $\log_2(p)$ increasing (slowly) without bound. Hence the minimum and maximum values for both *3EPO* and *CSPO* were expected to be found at smaller primes.

In fact these maxima and minima were found for very small primes: 5, 7, 11, 13 under the DR algorithm. In Table 9 the two maxima have been highlighted in green, two minima highlighted in red, giving the following deductions (assuming these maxima/minima hold over all integers):

- The series 1, 5, 25, 125, 625… has note labels C, E[5] = E', G#[$5^2$] = G#'', B#''', D##''''... and this sequence gives the fastest positive divergence of note labels from C over the integers
- The series 1, 7, 49, 343, 2401… has note labels C, Bb[7], Ab[$7^2$], Gb[$7^3$], Fb[$7^4$]… and is the fastest negative divergence of note labels from C over the integers
- The series 1, 11, 121, 1331… has rational commas [1], [11] = 33/32, [$11^2$] = $(33/32)^2$, [$11^3$] = $(33/32)^3$… and is the fastest upwards microtonal pitch shift of rational commas from 1/1, over integer commas [*n*/1]
- The series 1, 13, 169, 2197… has rational commas [1], [13] = 26/27, [$13^2$] = $(26/27)^2$, [$13^3$] = $(26/27)^3$… and is the fastest downwards microtonal pitch shift of rational commas from 1/1, over integer commas [*n*/1]

So 5 being in pitch class E[5] means $5^k$ gives the fastest divergence of note labels over the whole integers, when rational commas are allowed; however 3 being a G (see Table 4 earlier) means $3^k$ gives the fastest divergence of note labels over the Pythagorean integers, when rational commas are not allowed.

## 11) Shorthand notation for commas

The notation scheme can be tweaked for increased usability. For example, L[*x*/*y*]$_z$ is a specific note, and L[*x*/*y*] is a pitch class. Possible tweaks to make the pitch class quicker to write include:

- L[*x*] = L~*x* (the ~ symbol is much quicker to write than two square brackets [])
- L[1/*y*] = L_*y*
- L[*x*/*y*] = L~*x*_*y*
- L[5] = L' (it is especially useful to have shorthand for 5-limit notation)
- L[1/5] = L.
- L[25] = L''
- L[1/25] = L.. (and accordingly for more powers of 5)
- L[5*x*/*y*] = L'[*x*/*y*] (showing how a factor of 5 can be factored out into shorthand)
- L'. = L[5/5] = L (shorthands for 5 and 1/5 cancel out)
- L[7] = L~7
- L[1/7] = L_7 (shorthand makes writing undertones a lot quicker)
- 'p' and 'd' being used as shorthand for Pythagorean and inverse Pythagorean commas

These types of usability tweak can often be extended from pitch classes to notes, whenever octave number is recorded unambiguously, e.g. (L~*x*_*y*)$_z$ – however this particular form is not recommended since it requires more symbols than the original notation!

Shorthand notations such as L' and L. are similar in function to accidentals in other notations, such as the 5-limit accidentals in extended Helmholtz-Ellis (Sabat 2005). However, the translation problem increases as more accidentals are introduced. It seems optimal to have accidentals for [5] and [1/5], and possibly also for [7] and [1/7], but not for higher primes. For these it is unlikely to be worth the trade-off; increased ease of writing but decreased ease of translation, higher primes having a lower usage frequency.

Higher prime information can never be entirely replaced with accidentals, due to there being an infinite number of degrees of freedom to notate for, and a finite number of accidental definitions possible at any one time. Typically, each degree of freedom requires two accidentals to notate, one for the prime comma and one for its reciprocal. To access all higher primes there will be no alternative but to either list the higher primes (as in Kite's colour notation) or to encode them compactly into a notation similar to a rational number (as in versions of RCN presented here).

## 12) Application to pitch class lattices

One application for the shorthand notations is to write out pitch class lattices (Figure 3, Figure 4 below). These provide a geometric reference for common harmonies. For example, the C major triad (C E' G) becomes an upwards-pointing triangle in a pitch class lattice; the G minor triad (G Bb. D) becomes a downwards pointing triangle. Pitch class lattices provide a link between geometry and harmony. They are extremely useful for visualising harmonic spread and planning melodic progress in a composition, and for providing retuning options when converting a 12-EDO composition to JI.

Two useful pitch class lattices are the 5-limit (Figure 3) and 7-limit (Figure 4) variants. Horizontal movement corresponds to the pitch class changing by a power of 3; up-right and down-left diagonal movement to a power of 5; diagonally in or out of the page (Figure 4 only) a power of 7. Pythagorean series such as Bb, F, C, G, D, A, E, B, F# which vary by fifths/fourths are therefore oriented horizontally. These horizontal series reoccur at different positions in the lattice, each position corresponding to different 5-limit or 7-limit commas being applied to the series as a whole.

It is recommended that composers in Just Intonation use these types of pitch class lattice to give geometric insight into how their compositions travel around in 5-limit or 7-limit space. Often this helps suggest new or alternative notes, which would not be apparent without the pitch class lattice to visualise. It does not matter if a few notes in a composition have a higher $p$-limit; what is important is to understand the geometry of the 7-limit component of the music properly. The lattice is much more helpful with this than the prime factorisations or raw frequency information are. The lattice also provides a new way of looking at existing harmonies, for example, hearing an interesting song on the radio and tracking how its tune progresses through the different pitch classes, and whether comma pumps emerge, which are progressions such as C F D' G' C'.

Incidentally, JI comma pumps in notated music can be eliminated by writing music in locally tuned sections, finding section boundaries across which comma jumps occur, and stitching these sections together by retuning the reference frequency as a jump between the sections.

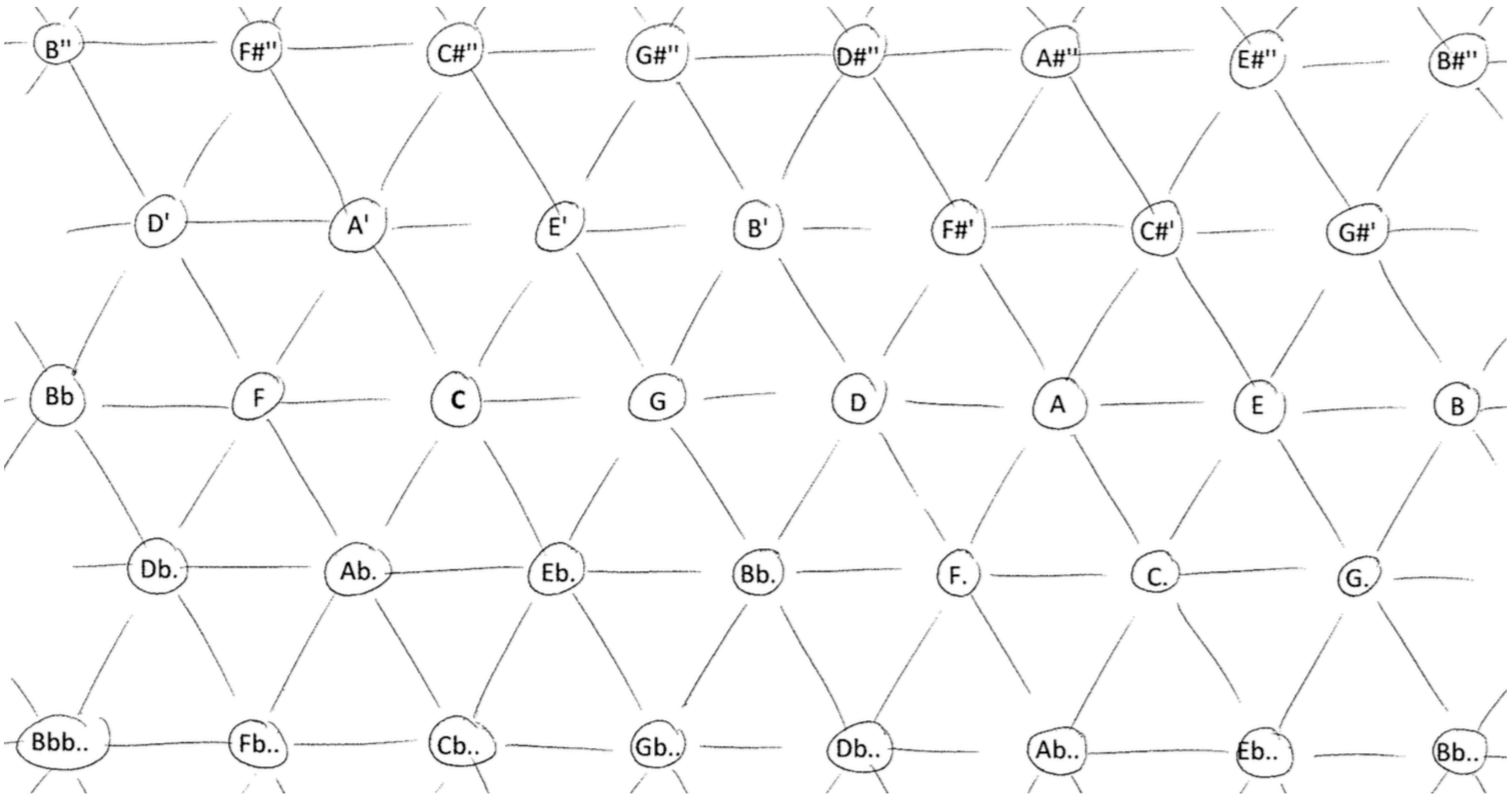

**Figure 3: 2D lattice for 5-limit pitch classes**

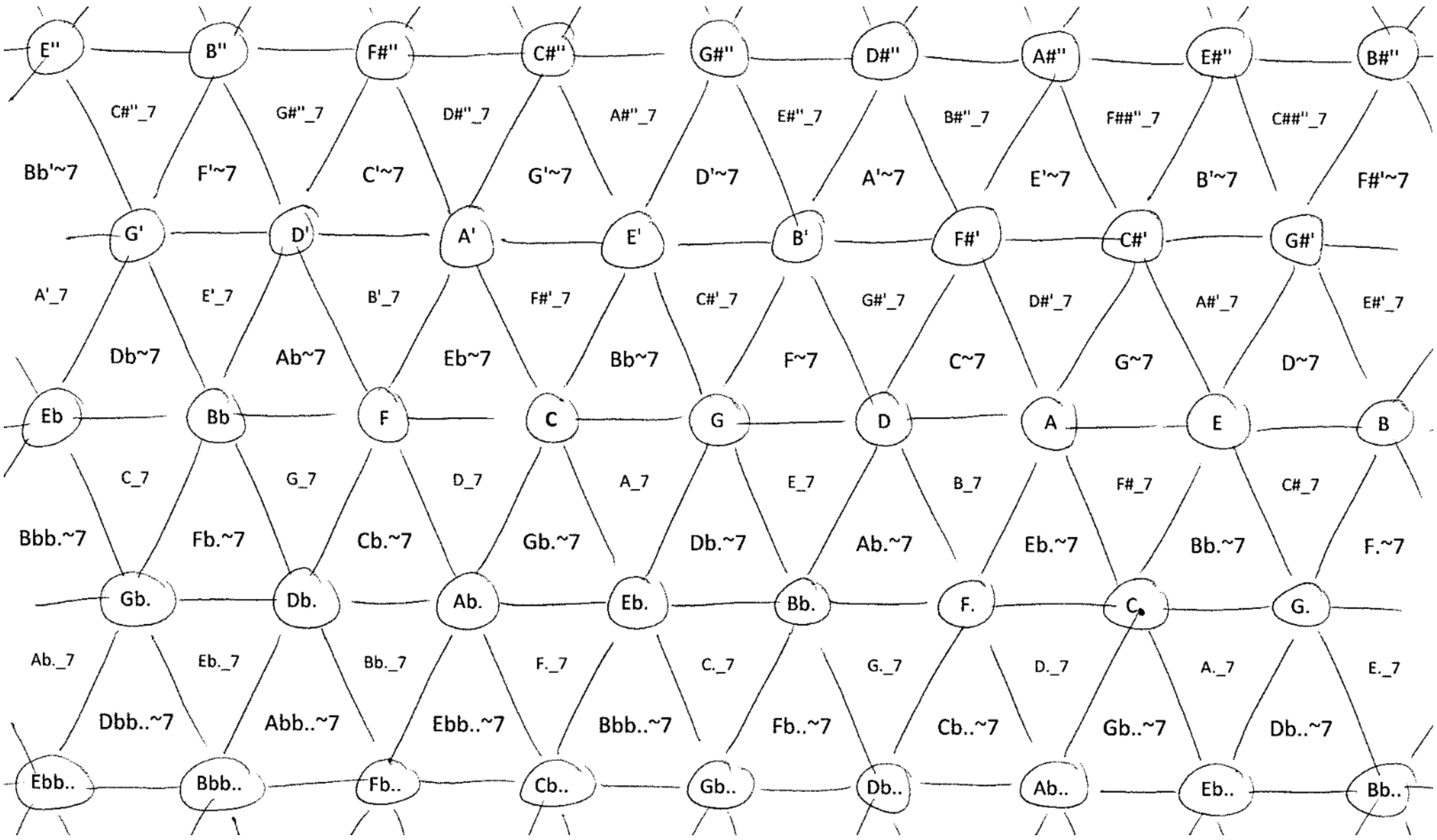

**Figure 4: Three layers of 3D lattice for 7-limit pitch classes**

Since this may cause the overall tuning to drift up or down during a piece, it is also possible to normalise out this tuning drift by moving the reference frequency slowly in the opposite direction. Carrying out this 'normalisation' smoothly in the 2 to 4 seconds around a comma jump seems to give the most satisfactory results, resulting in a level musical tuning which bends imperceptibly around comma jump sections. Einstein used the same technique in general relativity to describe gravity as a stitching together of local gravity maps, so normalised JI is like the ‘general relativity of musical tuning’, and can be recommended as a general technique for JI tuning applications.

## 13) Comparison of three algorithms (DR, SAG, KG2) for assigning 3-exponent *b*

The two algorithms SAG, KG2 from the literature are now compared with the novel algorithm DR presented in this paper. SAG is the scheme for Sagittal notation (Keenan & Taylor 2016) and KG2 is the (second proposed) scheme for Kite Giedraitis’s colour notation (Giedraitis 2017), both of these use 3-exponents from -6 to +6.

The Sagittal (SAG) algorithm:

- searches values of *b* in this order: 0, +/-1, +/-2, +/-3, +/-4, +/-5, +/-6
- accepts a comma if its size is below 68.57 cents. This value is half of a Pythagorean comma $B\#_3$ = 531441/524288 plus a $C\#_4$ = 2187/2048, i.e. smaller than $(3^{19}2^{-30})^{0.5}$
- if both + and – values of *b* give small enough candidates, choose *b* value of the smaller candidate comma
- Table 10 below can be used as a lookup table. Map *p* to a cents value in the octave, then find the range in Table 10 which includes this cents value. Choose the shared *b* value for this range.

**Table 10: Ranges in the octave for 3-exponent *b* in SAG algorithm**

| Start of Range | End of Range | Width of Range | *b* (SAG) | Pythag. Fraction | Pythag. Cents | Note Label | Octave Number |
|---|---|---|---|---|---|---|---|
| 0.00 | 68.57 | 68.57 | 0 | 1/1 | 0.00 | C | 4 |
| 68.57 | 135.34 | 66.76 | 5 | 256/243 | 90.22 | Db | 4 |
| 135.34 | 272.48 | 137.15 | -2 | 9/8 | 203.91 | D | 4 |
| 272.48 | 362.71 | 90.22 | 3 | 32/27 | 294.13 | Eb | 4 |
| 362.71 | 429.47 | 66.76 | -4 | 81/64 | 407.82 | E | 4 |
| 429.47 | 566.62 | 137.15 | 1 | 4/3 | 498.04 | F | 4 |
| 566.62 | 600.00 | 33.38 | 6 | 1024/729 | 588.27 | Gb | 4 |
| 600.00 | 633.38 | 33.38 | -6 | 729/512 | 611.73 | F# | 4 |
| 633.38 | 770.53 | 137.15 | -1 | 3/2 | 701.96 | G | 4 |
| 770.53 | 837.29 | 66.76 | 4 | 128/81 | 792.18 | Ab | 4 |
| 837.29 | 927.52 | 90.22 | -3 | 27/16 | 905.87 | A | 4 |
| 927.52 | 1064.66 | 137.15 | 2 | 16/9 | 996.09 | Bb | 4 |
| 1064.66 | 1131.43 | 66.76 | -5 | 243/128 | 1109.78 | B | 4 |
| 1131.43 | 1200.00 | 68.57 | 0 | 2/1 | 1200.00 | C | 5 |

In Table 10 the SAG algorithm has a maximum range width of 137.15 cents which occurs for 3-exponents -2, -1, 0, +1, +2 (summing the widths of the two occurrences of 0). Other 3-exponents have lower widths, meaning that there is a preference for 3-exponents near zero, implemented by searching the smaller 3-exponents first and stopping if an adequate comma has been found. The note labels for the ranges only get out of sequence once (F, Gb, F#, G) and this could be eliminated by combining 3-exponents 6 and -6 into one range of width 66.76 cents, perhaps using $b$ = -6 and F# due to its less complex commas with $p/2^n$.

Note that the last range has a note label in octave 5, $C_5$ = 2/1, however the commas produced from the $C_5$ range are still microtonal; for example, prime 127 maps to 127/64 in the octave, which is 1186.42 cents. The comma is thus (127/64)/(2/1) = 127/128 at -13.58 cents.

The Kite (KG2) algorithm for Kite Giedraitis's colour notation:

- maps prime numbers into the octave as a number of cents between 0 and 1200
- has 14 distinct ranges in the octave, each with a 3-exponent $b$ and note label in Table 11 below
- each range is either 50 or 100 cents wide
- the four ranges of width 50 cents are: 0-50, 550-600, 600-650, 1150-1200
- the rest of the octave is split into 100 cent ranges surrounding a 12-EDO note such as 50-150, 150-250 cents, etc.

**Table 11: Ranges in the octave for 3-exponent *b* in KG2 algorithm**

| Start of Range | End of Range | Width of range | 12 EDO | *b* (KG2) | Pythag. Fraction | Pythag. Cents | Interval Label | Note Label | Octave Number |
|---|---|---|---|---|---|---|---|---|---|
| 0 | 50 | 50 | 0 | 0 | 1/1 | 0.00 | P1 | C | 4 |
| 50 | 150 | 100 | 1 | 5 | 256/243 | 90.22 | m2 | Db | 4 |
| 150 | 250 | 100 | 2 | -2 | 9/8 | 203.91 | M2 | D | 4 |
| 250 | 350 | 100 | 3 | 3 | 32/27 | 294.13 | m3 | Eb | 4 |
| 350 | 450 | 100 | 4 | -4 | 81/64 | 407.82 | M3 | E | 4 |
| 450 | 550 | 100 | 5 | 1 | 4/3 | 498.04 | P4 | F | 4 |
| 550 | 600 | 50 | 6 | -6 | 729/512 | 611.73 | A4 | F# | 4 |
| 600 | 650 | 50 | 6 | 6 | 1024/729 | 588.27 | d5 | Gb | 4 |
| 650 | 750 | 100 | 7 | -1 | 3/2 | 701.96 | P5 | G | 4 |
| 750 | 850 | 100 | 8 | 4 | 128/81 | 792.18 | m6 | Ab | 4 |
| 850 | 950 | 100 | 9 | -3 | 27/16 | 905.87 | M6 | A | 4 |
| 950 | 1050 | 100 | 10 | 2 | 16/9 | 996.09 | m7 | Bb | 4 |
| 1050 | 1150 | 100 | 11 | -5 | 243/128 | 1109.78 | M7 | B | 4 |
| 1150 | 1200 | 50 | 12 | 0 | 2/1 | 1200.00 | P8 | C | 5 |

In Table 11 the KG2 algorithm has interval labels which use a descriptive letter: P for 'perfect, M for 'major', m for 'minor', d for 'diminished', A for 'augmented'; these directly relate to the note label used, e.g. 'G' (relative to C) for a perfect fifth 'P5', 'Eb' (relative to C) for a minor third 'm3'. This algorithm

KG2 keeps note labels in order, in particular the sequence F, F#, Gb, G; however it does so at the expense of fractions being out of order, since a Pythagorean F# (729/512, 611.73 cents) represents a range of 550 to 600 cents and comes before a Pythagorean Gb (1024/729, 588.27 cents) which represents 600 to 650 cents. This ordering aspect of algorithm KG2 is counterintuitive.

Now that algorithms SAG and KG2 have been presented, results are compared between the three algorithms DR, SAG and KG2:

**Table 12: Values of *b*, prime commas and note labels for algorithms DR, SAG and KG2**

| $p$ | $b_{DR}$ | $comma_{DR}$ | $label_{DR}$ | $b_{SAG}$ | $comma_{SAG}$ | $label_{SAG}$ | $b_{KG2}$ | $comma_{KG2}$ | $label_{KG2}$ |
|---|---|---|---|---|---|---|---|---|---|
| 5 | -4 | 80/81 | E | -4 | 80/81 | E | -4 | 80/81 | E |
| 7 | 2 | 63/64 | Bb | 2 | 63/64 | Bb | 2 | 63/64 | Bb |
| 11 | 1 | 33/32 | F | 1 | 33/32 | F | -6 | 704/729 | F# |
| 13 | -3 | 26/27 | A | -3 | 26/27 | A | 4 | 1053/1024 | Ab |
| 17 | -7 | 2176/2187 | C# | 5 | 4131/4096 | Db | 5 | 4131/4096 | Db |
| 19 | 3 | 513/512 | Eb | 3 | 513/512 | Eb | 3 | 513/512 | Eb |
| 23 | -6 | 736/729 | F# | -6 | 736/729 | F# | 6 | 16767/16384 | Gb |
| 29 | 2 | 261/256 | Bb | 2 | 261/256 | Bb | 2 | 261/256 | Bb |
| 31 | 0 | 31/32 | C | 0 | 31/32 | C | -5 | 248/243 | B |
| 37 | -2 | 37/36 | D | -2 | 37/36 | D | 3 | 999/1024 | Eb |
| 41 | -4 | 82/81 | E | -4 | 82/81 | E | -4 | 82/81 | E |
| 43 | 1 | 129/128 | F | 1 | 129/128 | F | 1 | 129/128 | F |
| 47 | -1 | 47/48 | G | -1 | 47/48 | G | -1 | 47/48 | G |
| 53 | -3 | 53/54 | A | -3 | 53/54 | A | -3 | 53/54 | A |
| 59 | -5 | 236/243 | B | 2 | 531/512 | Bb | -5 | 236/243 | B |
| 61 | -5 | 244/243 | B | -5 | 244/243 | B | -5 | 244/243 | B |
| 67 | -7 | 2144/2187 | C# | 5 | 16281/16384 | Db | 5 | 16281/16384 | Db |
| 71 | -2 | 71/72 | D | -2 | 71/72 | D | -2 | 71/72 | D |
| 73 | -2 | 73/72 | D | -2 | 73/72 | D | -2 | 73/72 | D |
| 79 | -4 | 79/81 | E | -4 | 79/81 | E | -4 | 79/81 | E |
| 83 | -4 | 83/81 | E | 1 | 249/256 | F | 1 | 249/256 | F |
| 89 | -6 | 712/729 | F# | 6 | 64881/65536 | Gb | -6 | 712/729 | F# |
| 97 | -1 | 97/96 | G | -1 | 97/96 | G | -1 | 97/96 | G |

In Table 12 a comparison has been made between 3-exponents from DR, SAG and KG2 algorithms, for primes from 5 to 97. Columns have been shaded to separate the different algorithms. The DR and SAG algorithms agree on all these primes except for 17, 59, 67, 83 and 89. The KG2 algorithm gives different results for some primes, including some slightly more complex commas: e.g. 16767/16384 for 23, making it a Gb; rather than 736/729 which makes 23 an F# in DR and SAG.

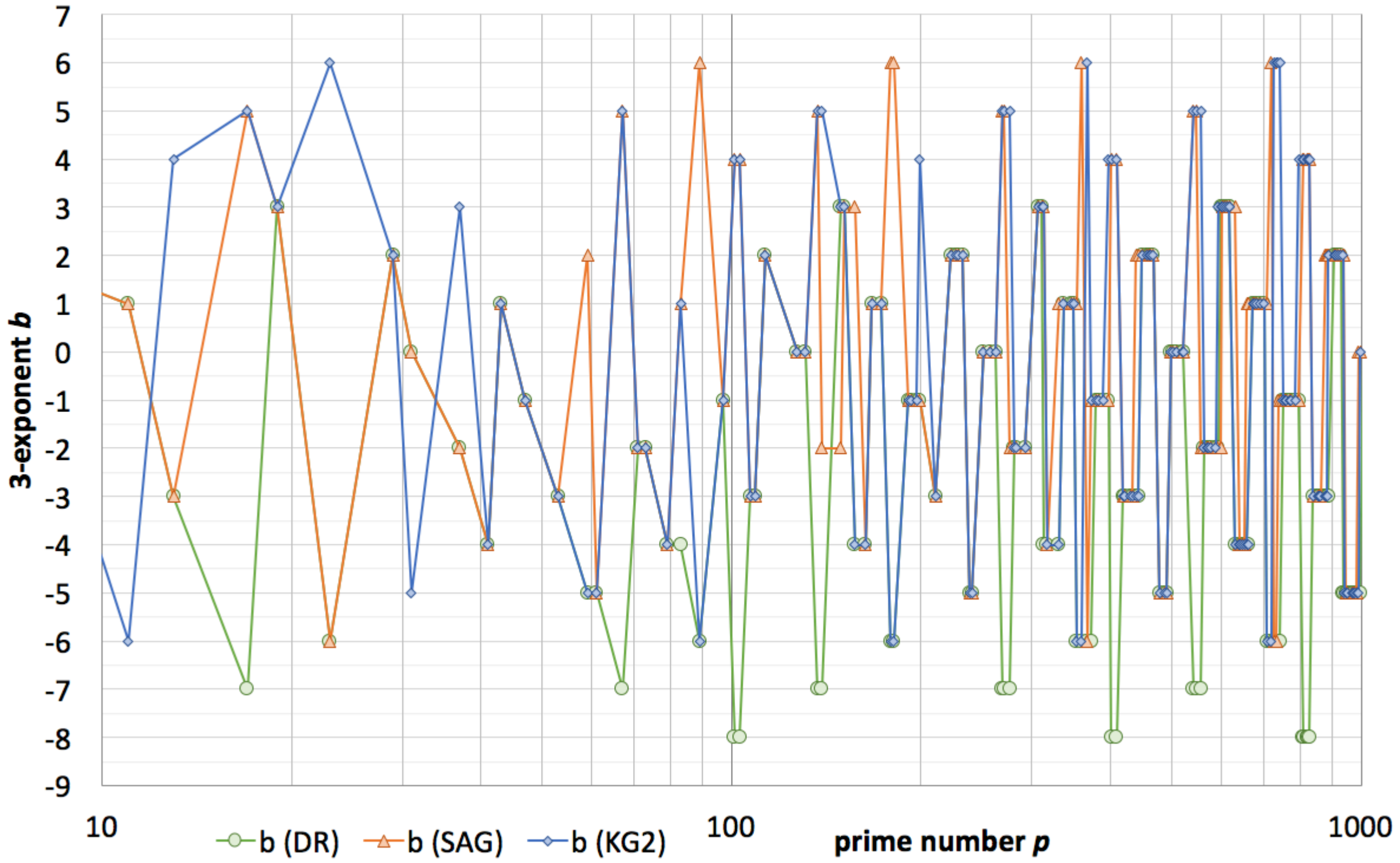

**Figure 5: Graph of 3-exponent *b* vs *p* for algorithms DR, SAG, KG2 (primes from 10 to 1000)**

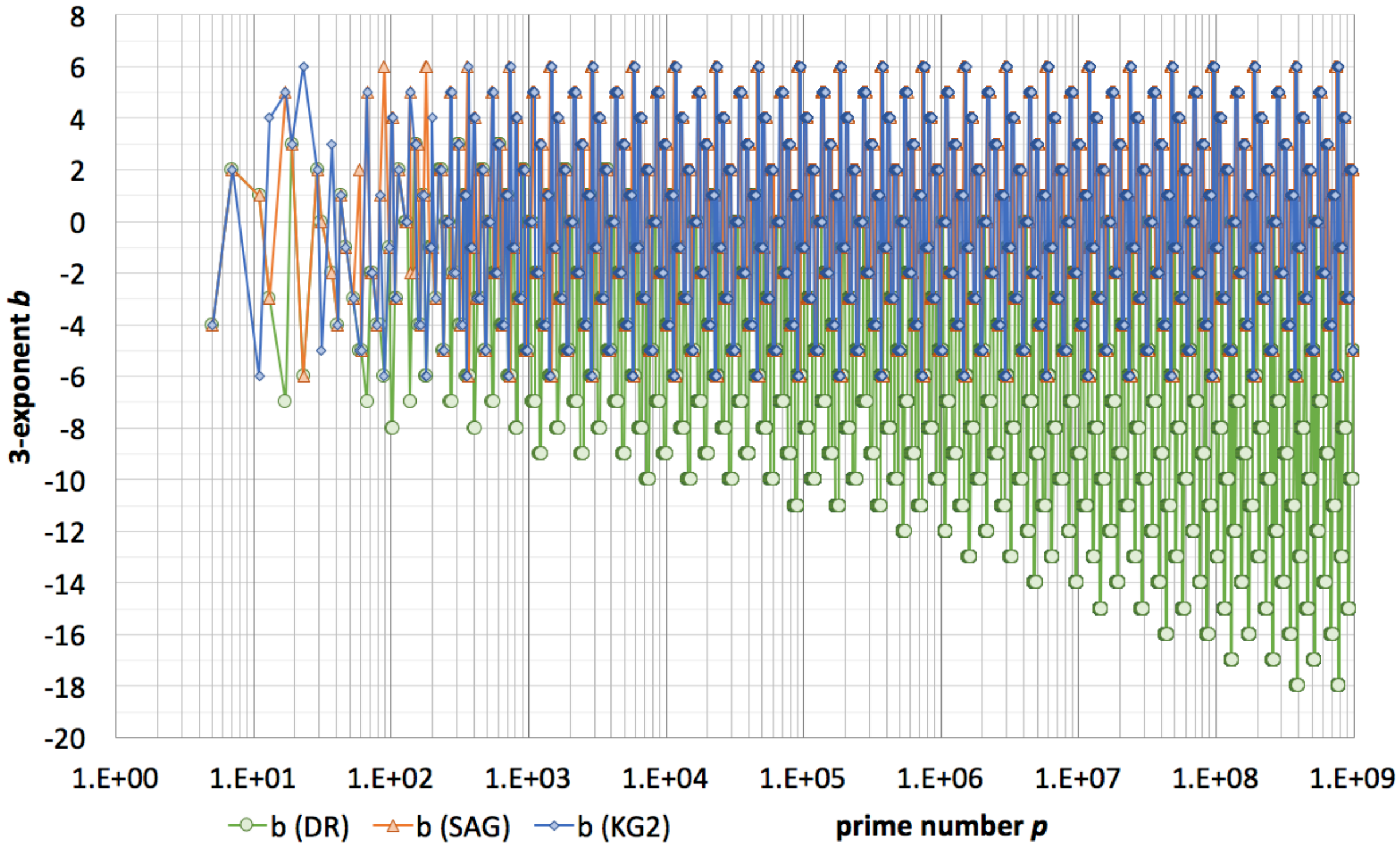

**Figure 6: Graph of *b* vs *p* for algorithms DR, SAG, KG2 (primes from 5 to $10^9$; data is partial above $10^5$)**

In Figure 5 the 3-exponents ($b$) are plotted for all three algorithms, for primes between 10 and 1000. The series coincide wherever the algorithms agree. This happens for primes 5, 7, 19, 29, 43… Otherwise for a specific prime there may be two or even three different values of $b$ for the different algorithms. The first prime to give a different result in all three algorithms is 139, with *b* values of -7 (DR), -2 (SAG) and +5 (KG2). For the three algorithms compared, it is expected that three different values are obtained for an unbounded subset of primes.

In Figure 6 the previous analysis is repeated for a much wider range of primes up to 1 billion. When the primes are small, the series are seemingly random; however by the time the primes have passed 2000 a pattern has emerged in the data series for all three algorithms; as *p* passes through each octave the *b* values cycle through the set of allowed integer values in a regular pattern.

The data plotted in Figure 6 was complete for primes from 1 to $10^5$, but partial for $10^5$ to $10^9$. The regular pattern for primes above 2000 was not interrupted by the partial data set; the regular cycle occurred once per octave; enough data points were included (over 1000 per octave) to produce every value of *b* in each cycle for each algorithm, which was possible since nearby primes share the value of *b*. So the regular pattern was seen to continue from $10^4$ to $10^9$ when using a logarithmic horizontal axis for the primes.

SAG and KG2 both used fixed ranges of *b* from +6 down to -6. For many primes the *b* values (and thus note names) agreed, however for some they did not, for example if SAG algorithm gave a Gb or F#, then KG2 always gave an F# or Gb respectively, a fact which could be deduced from cross-referencing the relevant ranges in the two lookup tables above (Table 10, Table 11).

DR (green) used a dynamic range for *b* which gradually became more negative in roughly linear relationship with $\log_3(p)$. Thus the three data series separate out for higher primes and no longer would be expected to coincide very much. Nonetheless, since $b = 0$ is allowed in algorithm DR for arbitrarily high *p*, the set of *p* for which all three algorithms agree is expected to be unbounded, for example, any primes near to $2^k$ for positive *k*.

In Figure 7 an empirical frequency distribution comparison is given for values of *b* from all three algorithms. The frequencies of *b* were counted under all three algorithms (DR, SAG, KG2) for primes from 50,000 to 100,000 (one octave). Since the prime numbers are not evenly distributed, a weighting of $\ln(p)/p$ was used in the counting process. This is based on the Prime Number Theorem which states the approximate count of primes (from 1 to *p*) is $p/\ln(p)$, so incorporating this weighting factor compensated for the natural tendency for higher cents ranges of an octave to contain more primes. A vertical axis has not been supplied since the weighted sums should be interpreted as relative measures. The vertical heights could however be scaled to make the area under each curve sum to 1. In that case, each curve would represent a probability distribution of *b* for each algorithm, for primes between 50,000 and 100,000.

DR algorithm gave negatively skewed values of *b*, but SAG and KG2 algorithms gave distributions symmetric about $b = 0$. DR gave an approximately uniform distribution for twelve values of *b*; KG2 gave the same uniform probability for eleven out of thirteen of its possible *b* values, with the remaining two *b* values (-6, +6) at half this probability. However, SAG had probability concentrated on the central five *b* values (-2, -1, 0, 1, 2), with probability tailing off for higher $|b|$. So *b* values were closest to 0 (on average) for SAG algorithm.

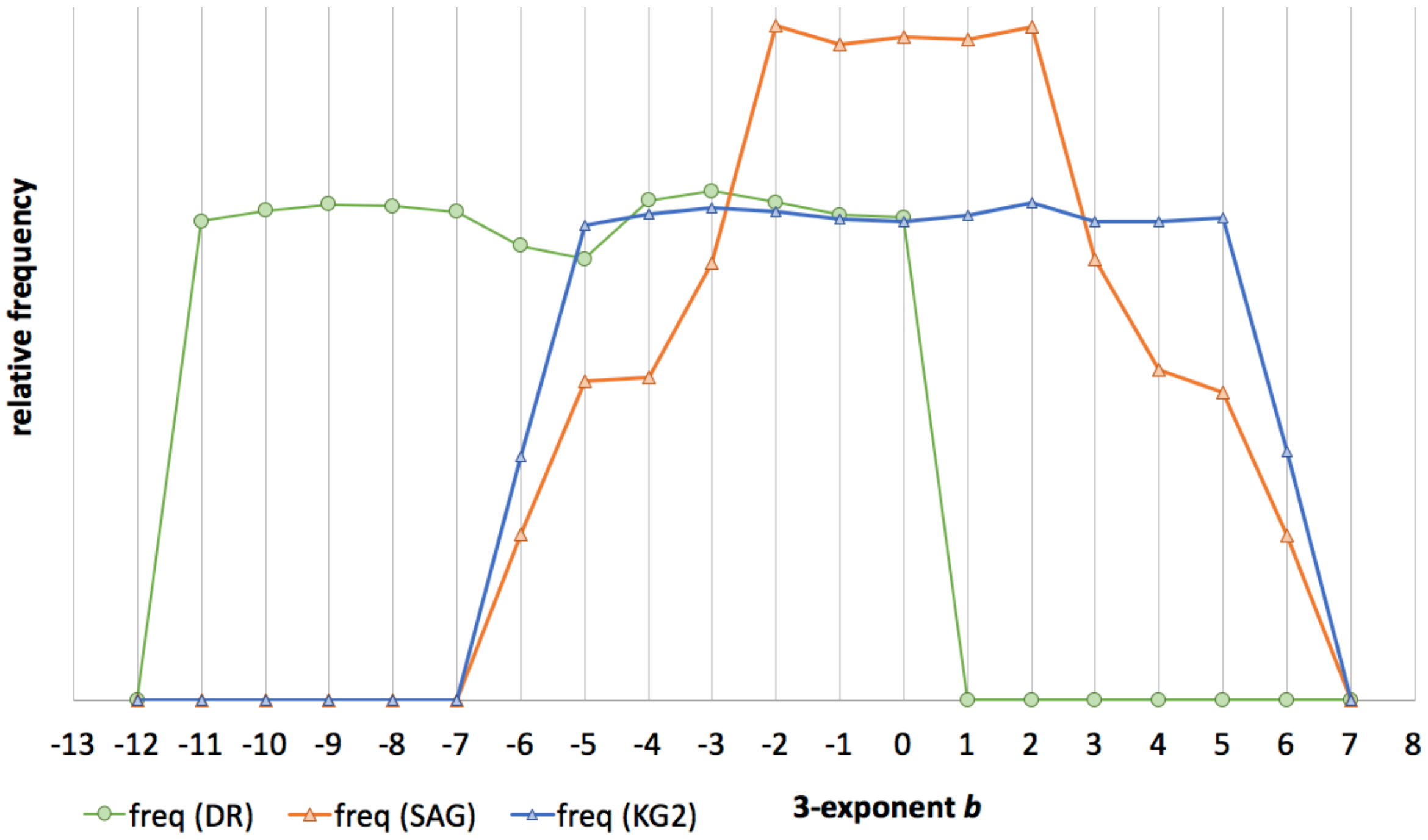

**Figure 7: Graph of relative frequency of each 3-exponent for primes from 50,000 to 100,000, weighted by ln(*p*)/*p*, for algorithms DR, SAG, KG2 (vertical axis omitted since heights are relative)**

There was a good agreement (for each *b*) between SAG and KG2 empirical probabilities in Figure 7 and the range widths given in the lookup tables Table 10, Table 11. This was expected since the larger the range width, the more likely a 'random' prime is to fall on that range. Hence this agreement helped give further confidence that algorithms SAG and KG2 had been implemented correctly.

If a different octave range of primes had been selected then SAG and KG2 algorithms should produce similar statistics, however DR algorithm would produce different statistics. Earlier in Table 7 it was stated that for DR algorithm the last $b = +1$ value is at $p = 45077$, and the first $b = -12$ value is at $p = 527869$. For octave ranges within these limits, the DR algorithm should produce approximately uniform distributions of *b* values, however this uniformity is not expected to hold outside of those limits.

When choosing which of these three algorithms to employ for assigning musical prime commas [*p*], there is a subjective aspect depending on personal preference. SAG tries to get the 3-exponent as low as possible. KG2 makes each prime fit in with nearby 12-EDO notes. DR tries to name primes based on existing names of nearby Pythagoreans, which is equivalent to minimising both comma size and comma complexity. The debate continues as to which approach is best.

It is the author's opinion that the latter process (DR) of naming primes according to nearby Pythagoreans is the right approach. Take the power of three $59049 = 3^{10}$ and the prime $p = 59051 = 3^{10}+2$. Under DR, SAG, KG2 algorithms, 59049 is always a Pythagorean A#, but 59051 is an A#, Bb, Bb respectively. Now 59091/59049 is 0.06 cents, and it is logical for primes at such small distances from 59049 to be notated as small modifications of its label, A#. On the other hand, Bb is 23 cents away and has no special relationship with 59051, like 59049 does. So surely 59051 must be an A#. (Although – with the shorthand 'p' for Pythagorean comma, A# = Bbp, meaning perhaps A# and Bb are not so different!) In any case, here are a series of these types of examples:

- 59051 (prime) = $3^{10}+2$ ought to be an A# (like 59049 = $3^{10}$), not a Bb as per SAG, KG2
- 19681 (prime) = $3^9–2$ ought to be a D# not an Eb as per SAG, KG2
- 6563 (prime) = $3^8+2$ ought to be a G# not an Ab as per SAG, KG2
- 2179 (prime) = $3^7–7$ ought to be a C# not a Db as per SAG, KG2
- 727 (prime) = $3^6–2$ ought to be an F# not a Gb as per KG2 (SAG gets this right)
- 241 (prime) = $3^5–2$ ought to be a B. All three algorithms assign this correctly.

For the first four examples, both SAG and KG2 assign the flatted note label instead of the sharped label. However, the Pythagorean series $3^n$ has note labels C, G, D, A, E, B, F#, C# (when starting at C = 1/1) and uses only sharped labels, not flatted labels. So it makes little sense to assign a prime near $3^n$ a flatted note label. The same reasoning holds for a prime near $2^m3^n$. This helps justify a prime naming convention based on nearby whole-numbered Pythagoreans, which implies prioritizing commas $2^a3^bp$ with $p$ in the numerator, implying the DR algorithm above. It also implies simpler commas, which are easier to work with, especially when translating between notation and frequency. Conversely, a criticism of SAG and KG2 algorithms is that they give too many flatted labels (which do not fit the primes), as can be seen in Figure 7 for the range $b > 0$, resulting in some unnecessarily complex commas in Table 12.

## 14) Translation between different free-JI algorithms and notations

Rational Comma Notation (RCN) can be used to notate the whole of free-JI; a specific version $RCN_X$ requires an algorithm X to be specified which assigns microtonal frequency adjustments [$p$] to each higher prime $p$. Three algorithms (DR, SAG, KG2) are available to make these assignments. Some frequencies have different RCN notations under different choices of algorithm, e.g. 11/8 = $F[11]_4$ under $RCN_{DR}$ and $RCN_{SAG}$, but 11/8 = $F\#[11]_4$ under $RCN_{KG2}$.

It is always possible to translate between any two versions of RCN. This is because each $RCN_X$ notation represents a rational number, and every rational number has an $RCN_X$ notation under any algorithm X. So to translate between $RCN_X$ and $RCN_Y$, first map notations in $RCN_X$ to numeric frequencies, and then map frequencies to notations in $RCN_Y$. So by using frequencies as an intermediate stage, all versions of RCN have translations between them.

In addition, SAG and KG2 algorithms have their own native plaintext formats for notation; Sagittal (SAG) in terms of (ASCII) arrows, and Kite's colour notation (KG2) in terms of colours and their acronyms. There will be correspondences between SAG plaintext and $RCN_{SAG}$ via numeric frequencies, and between KG2 plaintext and $RCN_{KG2}$ via numeric frequencies. So all the free-JI notation systems are of equal power, and translation is in theory possible between any two of them.

Since it is expected that different musicians will prefer different free-JI algorithms and/or notation schemes, it will be useful to have software which can automatically translate between these. Rather than try to specify this software in detail here, a few examples are given of translations; primarily in terms of versions of RCN, with optional translation into SAG or KG2 native plaintext:

<u>Example 1</u>

- Translate $E[5]_4$ and $Bb[7]_4$ from $RCN_{DR}$ into both $RCN_{SAG}$ and $RCN_{KG2}$

- Answer: The commas [5] and [7] are the same in all three algorithms, so $E[5]_4$ and $Bb[7]_4$ remain unchanged. In Kite's colour notation (KG2) native plaintext, E[5] is 'yellow-3rd' or 'y3', Bb[7] is 'blue-7th' or 'b7'

Example 2

- Translate $F[11]_4$ from $RCN_{DR}$ into $RCN_{KG2}$
- Answer: $F[11]_4$ is 11/8 with DR commas. 11/8 is $F\#[11]_4$ with KG2 commas
  In KG2 native plaintext 11/8 is 'j4' for 'jade-4th'

Example 3

- Translate $Db[17]_5$ from $RCN_{SAG}$ into $RCN_{DR}$
- Answer: $Db[17]_5$ is 17/8 with SAG commas. 17/8 is $C\#[17]_5$ with DR commas.

These are simple examples, but demonstrate the principle that music written in any particular free-JI notation could be translated into any other free-JI notation.

## 15) Conclusions

Rational Comma Notation (RCN) can notate the whole of free-JI, once an algorithm X has been chosen to give a specific version $RCN_X$ and make the assignments of prime commas [*p*] to all primes $p \geq 5$. $RCN_X$ gives every rational frequency in Just Intonation a notation of the form $L[x/y]_z$ where L is a Pythagorean (3-limit) pitch class and note label (e.g. C, E, Bb, G#, Dbbb, F######...), *z* is an octave number, and [*x*/*y*] is a rational comma for a small microtonal frequency adjustment using 5-rough numbers *x* and *y*, built up from prime commas [*p*]. The notation subset of the form $L_z$ is modified from Scientific Pitch Notation and notates all the 3-limit Pythagorean JI frequencies. Labels can also include optional shorthand notations 'p' and 'd' to simplify some note labelling and octave numbering using Pythagorean commas.

The novel algorithm DR set out above defines and calculates the microtonal fractions [*p*] for $p \geq 5$. It was compared to two other algorithms from the literature (SAG, KG2) and some advantages for each algorithm were stated. There is a subjective aspect to the design choices which affect the final choice of algorithm. For example, a design choice of what is most desirable: lower 3-exponents, note labels to be in a certain order, or primes to be named similarly to nearby integer Pythagoreans. In any case, the three algorithms (DR, SAG, KG2) agree on 7-limit commas, and two algorithms (DR, SAG) agree on commas to 53-limit, excluding 17. Translation is possible between any two free-JI notation schemes, so notation could potentially be automatically adapted to the user's preferred language.

Whichever algorithm X is used to define prime commas, once one has been settled on, it defines prime commas for all prime numbers, and these definitions allow all rational commas [*x*/*y*] to be constructed by multiplication of prime commas. Small notational tweaks (L~p  L'  L.  etc) allow useful shorthand for some subsets of RNC, in particular for pitch classes. Lattices provide a useful visualisation of pitch classes and are useful as a JI compositional aid or analytical device.

All the higher prime information is encoded into a rational comma of the form [*x*/*y*]. This uses a rational number as a shorthand for a frequency shift, itself a (usually different) rational number. Hence musicians would need a level of mathematical awareness in order to make full use of these [*x*/*y*] commas and access the higher primes, away from any potential 5-limit or 7-limit shorthand. No doubt there exist many musicians who are unwilling to cross over into mathematical territories, just as there are some mathematicians who cannot pick up a violin. However, the musical field is based on sound, vibrations, frequency and amplitude; these are all mathematically describable phenomena, so it is appropriate for musicians to have at least some basic level of mathematical understanding. It is likely that 12-EDO oversimplifies and hides the mathematical dimension of frequency from the musician. JI reveals this dimension again, and in the author's opinion we are better off for it.

By combining the disciplines of mathematics and music, there are rich pickings for the willing cross-disciplinarian. Such a musician may be truly able to navigate, understand, notate and compose in the rich infinite harmonic landscapes of Just Intonation. There are benefits to computer aided scoring and sequencing systems as well, in terms of presenting JI notes to the composer or musician which are more recognisable as musical pitches than a bare Hertz value or relative fraction would be. Starting with a major triad set of pitch classes (C E[5] G) or a minor triad (A C[1/5] E), a dominant seventh (G B' D F[7]) or a diminished chord (E. G Bb~7 C#~17) these commas can be used to produce the exact harmonic chords specified, to the benefit of all ears listening.

## 16) Nomenclature & Abbreviations

| | |
|---|---|
| *3EPO* | '3-Exponent Per Octave' – rate at which 3-exponent *b* diverges from 0, scaled by $\log_2(p)$ |
| *a* | The power of 2 in a prime comma |
| *AO* | 'Absolute Octaves' – absolute size of interval in octaves |
| *b* | The power of 3 in a prime comma |
| $b_{\text{mid}}$ | The midpoint for the range of *b* |
| $b_{\text{min, max}}$ | The minimum and maximum values of the range of *b* |
| *CSPO* | 'Comma Size Per Octave' – rate at which [*p*] in cents diverges from 0, scaled by $\log_2(p)$ |
| JI | Just Intonation |
| *LCY* | 'Log Complexity' – for 2 frequencies, same as Tenney Height |
| cents | 1 cent is an interval 1/100 of a semitone, or 1/1200 of an octave |
| *CM* | 'Comma Measure' – product of *AO* and *LCY* |
| DR | Prime comma algorithm presented in this paper |
| EDO | Equal Division of the Octave, e.g. 7-EDO divides the octave into 7 equal intervals |
| 12-EDO | Equal Division of the Octave into 12 semitones. Currently widely used as 'the' tuning. |
| JI | Just Intonation – where all frequency ratios are 'just', comprising whole numbers |
| KG2 | The (second) prime comma algorithm originating from Kite Giedraitis (2017) and associated with Kite's colour notation |
| lattice | A geometric arrangement of notes or pitch classes with constant ratios in straight lines |
| octave | Frequency change of a factor of 2 |
| $p$ | A prime number |
| $[p]$ | Prime comma corresponding to the prime $p$ |
| pitch class | Frequencies which are a whole number of octaves apart |
| $p$-limit | Just Intonation restricted to rational frequencies which are $p$-smooth numbers |
| $p$-smooth | Rational numbers with all prime factors $p$ or smaller |
| $p$-rough | Rational numbers with all prime factors $p$ or larger (e.g. 5-rough numbers) |
| $P_B$ | Value of 375,787 which is the boundary where SR is a proper subset of PR |
| PR | Primary range of candidate commas with $p$ in the numerator |
| Pythagorean | 3-limit JI; frequencies of the form $2^a3^b$ |
| $RCN_X$ | Rational Comma Notation for free-JI; combining modified SPN with rational commas; X (if specified) is the algorithm used to assign prime commas $[p]$ for all higher primes $p$ |
| RI | Rational Intonation – some authors prefer using the term RI for high $p$-limits, and JI only when the $p$-limit is low, say below 20 or 50. |
| semitone | An interval 1/12 of an octave |
| SAG | Prime comma algorithm originating from a Sagittal forum (Keenan & Taylor 2016) |
| SPN | Scientific Pitch Notation; middle C is $C_4$, concert A is $A_4$, C above middle C is $C_5$ |
| SR | Secondary range of 12 candidate commas with $b$ value closest to $b_{mid}$ |
| X | Represents a choice of algorithm (e.g. DR, SAG, KG) |

## 18) Acknowledgements

Dave Keenan who offered constructive criticism on a previous algorithm which chose some suboptimal commas, in particular for primes 11 and 13. Kite Giedraitis for discussions of his colour notation and comparisons with this comma notation. Also, other members of the Xenharmonic Facebook groups who read my earlier drafts and made helpful comments, including: Juhani Nuorvala, Graham Breed, Andrew

Meronek, George Secor, Cam Taylor, William Lynch, Carl Lumma. The Xenharmonic groups generally for being willing to discuss many of the ideas in this paper through many discussion threads.

## 19) Author contact details

| | | |
|---|---|---|
| ORC ID | http://orcid.org/0000-0002-4785-9766 | Academic profile |
| arXiv | https://arxiv.org/a/ryan_d_1 | Papers (pre-prints) |
| SoundCloud | https://soundcloud.com/daveryan23/tracks | JI music examples |
| LinkedIn | https://www.linkedin.com/in/davidryan59 | Professional page |
| Email: | david ryan 1998 @ hotmail.com | (remove spaces) |